\documentclass[format=acmsmall, review=false, screen=true]{acmart}
\renewcommand\footnotetextcopyrightpermission[1]{} 
\pdfoutput=1

\usepackage[ruled]{algorithm2e}
\usepackage{amsmath}
\usepackage{subfig}
\usepackage{bm}
\usepackage{rotating}
\usepackage{array}
\newcolumntype{?}{!{\vrule width 1pt}}
\usepackage{tablefootnote}
\usepackage{makecell}
\usepackage{multirow}
\usepackage{forest}
\usetikzlibrary{arrows.meta}

\SetAlFnt{\small}
\SetAlCapFnt{\small}
\SetAlCapNameFnt{\small}
\SetAlCapHSkip{0pt}
\IncMargin{-\parindent}
\newcommand{\major}[1]{\textcolor{black}{#1}} 
\newcommand{\minor}[1]{\textcolor{black}{#1}}  

\newcommand{\indexI}{{i}}

\newcommand{\totalopt}{{n}}

\newcommand{\optname}{{o}}
\newcommand{\opt}{{\optname_{\indexI}}}
\newcommand{\optvector}{{\boldsymbol{\optname}}}
\newcommand{\optspace}{{\mathcal{O}}}

\newcommand{\optimalvector}{{\bar{\optvector}}}

\newcommand{\app}{{a}}
\newcommand{\appset}{{A}}
\newcommand{\appsetSize}{{m}}

\newcommand{\acnumber}{{l}}
\newcommand{\acname}{{\gamma}}

\newcommand{\acvector}{{\boldsymbol{\acname}}}

\newcommand{\profileMatrix}{{P}}

\newcommand{\rcname}{{\alpha}}

\newcommand{\rcvector}{{\boldsymbol{\rcname}}}

\newcommand{\pcaname}{{\rho}}

\newcommand{\pcavector}{{\boldsymbol{\pcaname}}}

\newcolumntype{M}[1]{>{\centering\arraybackslash}m{#1}}

\setcopyright{none}

\received{November 2016}
\received[revised]{August 2017}
\received[revised]{February 2018}
\received[accepted]{March 2018}

\begin{document}
\title[A Survey on Compiler Autotuning using Machine Learning]{A Survey on Compiler Autotuning using Machine Learning}  
\subtitle{\normalsize{Accepted in ACM Computing Surveys 2018 (\footnotesize{Received Nov 2016, Revised Aug 2017}) \\ https://doi.org/10.1145/3197978}}

\author{Amir H. Ashouri}
\orcid{0000-0001-8606-6497}
\affiliation{%
  \institution{University of Toronto, Canada}
  \streetaddress{10 King's College Rd}
  \city{Toronto}
  \state{Ontario}
  \postcode{M5S 3G8}
  \country{Canada}}
\author{William Killian}
\affiliation{%
  \institution{Millersville University of Pennsylvania, USA}
  \city{Millersville}
  \state{PA}
  \postcode{19551}
  \country{USA}
}
\author{John Cavazos}
\affiliation{%
  \institution{University of Delaware, USA}
  \department{Electrical and Computer Engineering}
  \city{Newark}
  \state{DE}
  \postcode{19716}
  \country{USA}
}
\author{Gianluca Palermo}
\affiliation{%
  \institution{Politecnico di Milano, Italy}
  \streetaddress{Via Ponzio, 34/5}
  \city{Milan}
  \state{Lombardia}
  \postcode{20133}
  \country{ITALY}
}
\author{Cristina Silvano}
\affiliation{%
  \institution{Politecnico di Milano, Italy}
  \streetaddress{Via Ponzio, 34/5}
  \city{Milan}
  \state{Lombardia}
  \postcode{20133}
  \country{ITALY}
}

\begin{abstract}
Since the mid-1990s, researchers have been trying to use machine-learning based approaches to solve a number of different compiler optimization problems.
These techniques primarily enhance the quality of the obtained results and, more importantly, make it feasible to tackle two main compiler optimization problems: optimization selection (choosing which optimizations to apply) and phase-ordering (choosing the order of applying optimizations).
The compiler optimization space continues to grow due to the advancement of applications, increasing number of compiler optimizations, and new target architectures.
Generic optimization passes in compilers cannot fully leverage newly introduced optimizations and, therefore, cannot keep up with the pace of increasing options.
This survey summarizes and classifies the recent advances in using machine learning for the compiler optimization field, particularly on the two major problems of (1) selecting the best optimizations, and (2) the phase-ordering of optimizations.
The survey highlights the approaches taken so far, the obtained results, the fine-grain classification among different approaches and finally, the influential papers of the field.
\end{abstract}

\keywords{Compilers, Autotuning, Machine Learning, Phase ordering, Optimizations, Application Characterization}

\thanks{This work is partially supported by the EU Commission 
        H2020-FET-HPC program under the grant ANTAREX-671623.
 Author's addresses: AH. Ashouri, ECE Department, University of Toronto, Canada; w. Killian {and} J. Cavazos, Department of Computer and Information Science, University of Delaware, USA;  G. Palermo {and} C. Silvano, DEIB, Politecnico di Milano, Italy; 
  }

\maketitle

\renewcommand{\shortauthors}{A. H. Ashouri et al.}

\section{Introduction}
\label{sec:introduction}

Moore's Law \cite{schaller1997moore} states that transistor density doubles every two years.
However, the rate at which compilers improve is on the order of a few percentage points each year.
Compilers are necessary tools bridging the gap between written software and target hardware.
There are many unsolved research challenges in the domain of compilers~\cite{Hall2009}.
Entering the \emph{post Moore's Law} era \cite{esmaeilzadeh2011dark}, compilers struggle to keep up with the increasing development pace of ever-expanding hardware landscape (e.g., CPUs, GPUs, FPGAs) and software programming paradigms and models (e.g., OpenMP, MPI, OpenCL, and OpenACC).
Additionally, the growing complexity of modern compilers and increasing concerns over security are among the most serious issues that the compilation research community faces. 

Compilers have been used for the past 50 years ~\cite{aho1986compilers,Hall2009} for generating machine-dependent executable binary from high-level programming languages.
Compiler developers typically design optimization passes to transform each code segment of a program to produce an optimized version of an application.
The optimizations can be applied at different stages of the compilation process since compilers have three main layers: (1) \emph{front-end} (2) \emph{intermediate-representation} (IR) and (3) \emph{backend}.
At the same time, optimizing source code by hand is a tedious task.
Compiler optimizations provide automatic methods for transforming code.
To this end, optimizing the intermediate phase plays an important role in the performance metrics. Enabling compiler optimization parameters (e.g., loop unrolling, register allocation, etc.) could substantially benefit several performance metrics.
Depending on the objectives, these metrics could be execution time, code size, or power consumption.
A holistic exploration approach to trade-off these metrics also represents a challenging problem~\cite{zaccaria2005most}. 

\emph{Autotuning} \cite{Vuduc2011,basu2013towards} addresses automatic code-generation and optimization by using different scenarios and architectures.
It constructs techniques for automatic optimization of different parameters to maximize or minimize the satisfiability of an objective function. Historically, several optimizations were done in the backend where \emph{scheduling}, \emph{resource-allocation} and \emph{code-generation} are done~\cite{chaitin1981register,register1992phaseOrdering}.
The constraints and resources form a linear system (ILP) which needs to be solved. Recently, researchers have shown increased effort in introducing front-end and IR-optimizations.
Two observations support this claim: (1) the complexity of a backend compiler requires exclusive knowledge strictly by the compiler designers, and (2) lower overheads with external compiler modification compared with back-end modifications.
The IR-optimization process normally involves fine-tuning compiler optimization parameters by a multi-objective optimization formulation which can be harder to explore. Nonetheless, each approach has its benefits and drawbacks and are subject to analysis under their scope. 

A major challenge in choosing the right set of compiler optimizations is the fact that these code optimizations are programming language, application, and architecture dependent.
Additionally, the word optimization is a misnomer --- there is no guarantee the transformed code will perform better than the original version.
In fact, aggressive optimizations can even degrade the performance of the code to which they are applied ~\cite{Triantafyllis2003}.
Understanding the behavior of the optimizations, the perceived effects on the source-code, and the interaction of the optimizations with each other are complex modeling problems.
This understanding is particularly difficult because compiler developers must consider hundreds of different optimizations that can be applied during the various compilation phases.
The phase-ordering problem is realized when considering the order which these hundreds of optimizations are used.
There are several open problems associated with the compiler optimization field that have not been adequately tackled \cite{zhao2005model}.
These problems include finding \emph{what} optimizations to use, \emph{which} set of parameters to choose from (e.g., loop tiling size, etc.), and in \emph{which order} to apply them to yield the best performance improvement.
The first two questions create the problem of \emph{selecting the best compiler optimizations} and the taking into account the third question is forming the \emph{phase-ordering} problem of compiler optimizations.

The problem of \emph{phase-ordering} has been an open problem in the field of compiler research for many decades~\cite{Vegdahl1982,Whitfield1990,Kulkarni2007,Kulkarni2012,Ashouri2017micomp}.
The inability of researchers to fully address the phase-ordering problem has led to advances in the simpler problem of selecting the right set of optimizations, but even the latter has yet to be solved ~\cite{bodin1998iterative,chen2012deconstructing,Ashouri2016Cobayn}.
The process of selecting the right optimizations for snippets of code, or \emph{kernels}, is typically done manually, and the sequence of optimizations is constructed with little insight into the interaction between the preceding compiler optimizations in the sequence.
The task of constructing heuristics to select the right sequence of compiler optimizations is infeasible given the increasing number of compiler optimizations being integrated into compiler frameworks.
For example, GCC  has more than 200 compiler passes, referred to as compiler \emph{options} \footnote{https://gcc.gnu.org/onlinedocs/gcc/Option-Summary.html}, and LLVM-clang and LLVM-opt each have more than 150 transformation \emph{passes} \footnote{http://llvm.org/docs/Passes.html}.
Additionally, these optimizations are targeted at different phases of the compilation process.
Some of these passes only analyze data access while others may look at loop nests.
Most compiler optimization flags are disabled by default, and compiler developers rely on software developers to know which optimizations should be beneficial for their application.
Compiler developers provide standard ``named'' optimization levels, e.g. \texttt{-O1, -O2, -Os}, etc. to introduce a fixed-sequence of compiler optimizations that on average achieve better performance on a set of benchmark applications chosen by the compiler developers.
However, using predefined optimizations usually is not good enough to bring the best achievable application-specific performance.
One of the key approaches recently used in the literature to find the best optimizations to apply given an application is inducing prediction models using different classes of machine learning~\cite{alpaydin2014introduction}.
The central focus of this survey is to highlight approaches which leverage machine learning to find the best optimizations to apply.

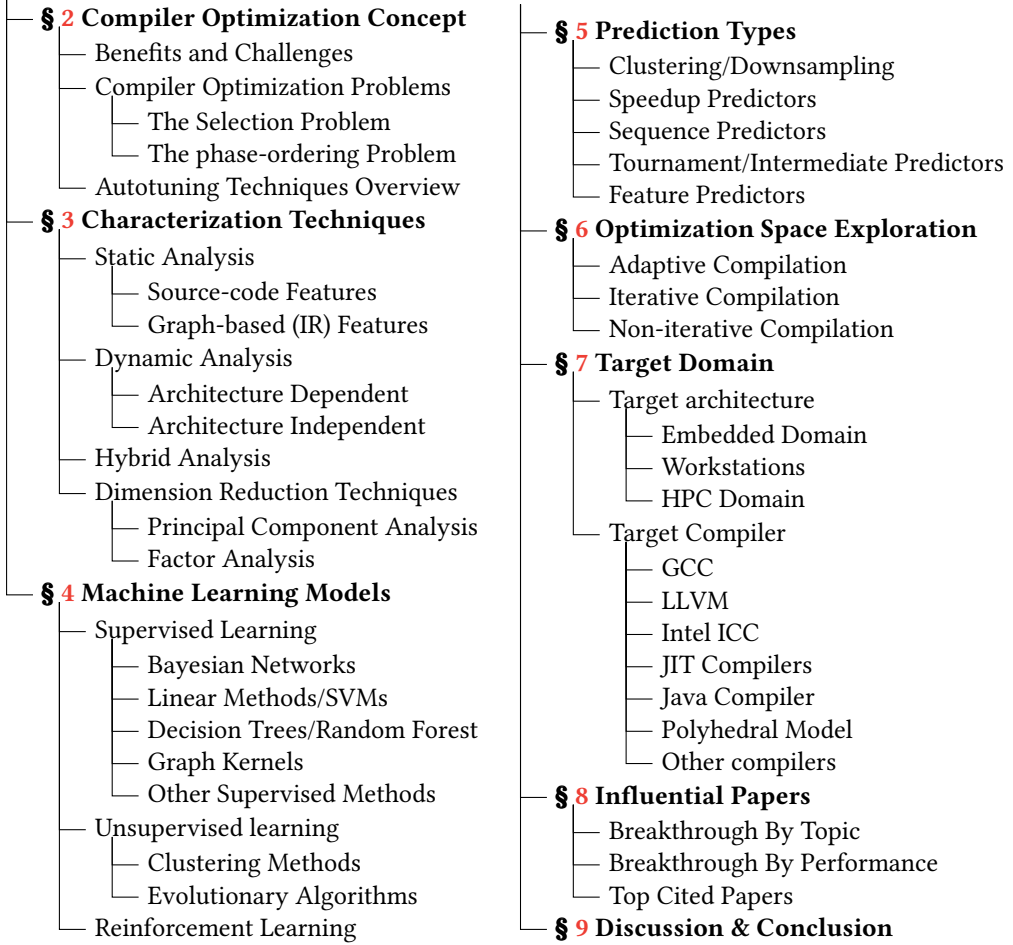
\begin{figure}[t!]
\centering
\subfloat{
\label{fig:contribution_left}
\begin{forest}
  dir tree
  [\textbf{\large{Survey Organization}} 
    [\textbf{\pmb{\S}~\ref{sec:promises and challenges} Compiler Optimization Concept}
		[Benefits and Challenges]
		[Compiler Optimization Problems
			[The Selection Problem]
			[The phase-ordering Problem]	
		]
        [\major{Autotuning Techniques Overview}]
    ]    
    [\textbf{\pmb{\S}~\ref{sec:characterization techniques} Characterization Techniques}
      [Static Analysis
       [Source-code Features]
       [Graph-based (IR) Features]
      ]
      [Dynamic Analysis
		[Architecture Dependent]
		[Architecture Independent]      
      ]
      [Hybrid Analysis]
      [Dimension Reduction Techniques
        [Principal Component Analysis]
        [Factor Analysis]
      ]
    ]
    [\textbf{\pmb{\S}~\ref{sec:ML_Models}  Machine Learning Models}
      [Supervised Learning
        [Bayesian Networks]
        [Linear Methods/SVMs]
        [Decision Trees/Random Forest]
        [Graph Kernels]
        [Other Supervised Methods]
      ]
      [Unsupervised learning
        [Clustering Methods]
        [Evolutionary Algorithms]
      ]
      [Reinforcement Learning]
    ]
  ]
\end{forest}}
\subfloat{
\label{fig:contribution_right}
\begin{forest}
  dir tree
  [
   [\textbf{\pmb{\S}~\ref{sec:predication_classes}  Prediction Types}
      [Clustering/Downsampling]
      [Speedup Predictors]
      [Sequence Predictors]
      [Tournament/Intermediate Predictors]
      [Feature Predictors]
   ]
   [\textbf{\pmb{\S}~\ref{sec:DSE} Optimization Space Exploration}
		[Adaptive Compilation]
		[Iterative Compilation]
		[Non-iterative Compilation]   
   ]
   [\textbf{\pmb{\S}~\ref{sec:Target Domain} Target Domain}
      [Target architecture
	      [Embedded Domain]
    	  [Workstations] 
    	  [HPC Domain]
      ]
      [Target Compiler 
    	  [GCC]
    	  [LLVM]
    	  [Intel ICC]
    	  [JIT Compilers]
    	  [Java Compiler]
    	  [Polyhedral Model]
    	  [Other compilers]
      ] 
    ] 
     [\textbf{\pmb{\S}~\ref{sec:influentials} Influential Papers}
		  [Breakthrough By Topic]
          [\major{Breakthrough By Performance}]
		  [Top Cited Papers]
     ]
     [\textbf{\pmb{\S}~\ref{sec:conclusion} Discussion \& Conclusion}]
  ]
\end{forest}}
\caption{Organization of the survey in different sections}
\label{fig:contribution}
\end{figure}

\label{sec:introduction:contribution}
\textbf{Contribution}.  
In this survey, we provide an extensive overview of techniques and approaches proposed by the authors tackling the aforementioned problems.
We highlight over 200 recent papers proposed for compiler autotuning when Machine Learning (ML) was used.
The classification in different subfields is done by several representative features shown in Figure \ref{fig:contribution}.
To the best of our knowledge, the first application of machine learning adapted for autotuning compilers were proposed by~\cite{Koseki1997,cavazos1998,Cooper1999}.
However, there were other original works which discuss the problems and acted as the driving force of using machine learning ~\cite{basil1975iterative,wulf1975design,loveman1977program,Vegdahl1982,padua1986advanced,pollock1990,click1995combining,Whitfield1997,bodin1998iterative}.
Thus, we consider the past 25 years of research related to autotuning compilers as it covers the entire time span of the literature \footnote{We look forward to keep updating the current survey on a quarterly basis at its arXiv branch \cite{ashouri2018surveyArXiv} and we encourage researchers to inform us about their new related works.}.
Additionally, this article can be leveraged as a connecting point for two existing surveys ~\cite{schneck1973survey,Bacon1994} on the compiler optimization field.

\label{sec:introduction:organization}
\textbf{Organization}.  
\major{We organize this survey as follows: We start by providing an overview of the different methods of data acquisition; followed by presenting preprocessing techniques.
We then discuss different machine learning models and the types of objectives a model is constructed.
Finally, we elaborate on the different target platforms these techniques have been applied.
Additionally, to facilitate this organizational flow depicted in Figure \ref{fig:contribution}, we provide Figure \ref{fig:sampleAutotuningFramework} for which we show a sample autotuning framework leveraging compilers and machine learning.
We discuss more on the autotuning nomenclatures in Section \ref{sec:introduction:optimizationSpace:autotuningTechniques}.}
Section~\ref{sec:promises and challenges} discusses the motivations and challenges involved in the compiler optimization research; this is followed by an analysis of the optimization space for the two major optimization problems in Section \ref{sec:introduction:optimizationSpace}. 
We review the existing characterization techniques and the classification of those in Section~\ref{sec:characterization techniques}.
Furthermore, we discuss the machine learning models used in Section~\ref{sec:ML_Models} and provide a full classification of different prediction techniques used in the recent research in Section~\ref{sec:predication_classes}.
Section~\ref{sec:DSE} examines the various space exploration techniques on how optimization configurations are traversed within the optimization space.
In Section~\ref{sec:Target Domain}, we discuss the different target architectures and compiler frameworks involved in the tuning process.
We include a brief review of the Polyhedral compilation framework along with other widely used compiler frameworks in Section~\ref{sec:Target Domain}.
\major{Section ~\ref{sec:influentials} includes a discussion on the influential papers of the field classified by their obtained (1) performance, (2) novelty, influence they had on the succeeding work, and (3) number of citations.}
Finally, we conclude the article with a brief discussion on past and future trends of compiler autotuning methodologies using machine learning.

\forestset{
  dir tree/.style={
    for tree={
      parent anchor=south west,
      child anchor=west,
      anchor=mid west,
      inner ysep=-3.5pt,
      grow'=0,
      align=left,
      edge path={
        \noexpand\path [draw, \forestoption{edge}] (!u.parent anchor) ++(1em,0) |- (.child anchor)\forestoption{edge label};
      },
      if n children=0{}{
        delay={
          prepend={[,phantom, calign with current]}
        }
      },
      fit=rectangle,
      before computing xy={
        l=2em
      }
    },
  }
}

\label{sec:introduction:scope}
\textbf{Scope of the survey}. 
We organize the research performed in this survey in different categories to highlight their similarities and differences.
However, it is important to note that the presented study has many potential classifications in the domain of machine learning and compilers research.
We organize the survey in a way that all research papers corresponding to a specific type of classification are cited under each classification.
\minor{Furthermore, under every classification,  we selectively picked the more notable works and we provide more elaboration on their contribution. As an example, Cavazos et al.~\cite{Cavazos2007} proposed to use performance counters to characterize an application.
The vector of features then can be used to construct prediction models.
Since this work was using this novel way of characterization, it is elaborated more on the Section~\ref{sec:characterization techniques:dynamic:dependent}.
Due to our classification policy, this work is also cited in other classification tables (prediction type, target platform)}.
We hope this survey will be useful for a wide range of readers including computer architects, compiler developers, researchers, and technical professionals.

\section{Compiler Optimizations} 
\label{sec:promises and challenges}

\subsection{A Note on Terminology and Metrics}
\label{sec:introduction:terminology}

Since research works mentioned in this survey originated from varying authors, terminologies may be locally and contextually defined. This results in term definitions which may not strictly be defined to apply to all cited publications.
We clarify terms used in this survey here and relate them to the publications discussed.
The field of compiler optimization has been referred to as \emph{compiler autotuning}, \emph{compilation}, \emph{code optimization}, \emph{collective optimization}, and \emph{program optimization}.
To maintain clarity, we do not use all these terms but instead use \emph{optimizing compilers}, or \emph{compiler autotuning}.
Moreover, under each classified subsection, we will point out the other nomenclatures that have been used widely and our reference subsequently.

\subsection{Compiler Optimization Challenges and Benefits}
\label{sec:introduction:benefits}

The majority of potential speedup no longer arrives at the increase of processor core clock frequencies.
Automatic methods of exploiting parallelism and reducing dependencies are needed. As such,
compiler optimizations \cite{padua1986advanced} allow a user to affect the generated code without changing the original high-level source code.
When these optimizations are applied may result in a program running better on a target architecture.
Since a user is not able to manually tune a large code base, automatic methods must be introduced.
Furthermore, manual tuning is not portable -- transformations applied to code running on one architecture is not guaranteed to yield the same performance increase on another architecture.

\subsection{Compiler Optimization Problems}
\label{sec:introduction:optimizationSpace}

The problem of interdependency among phases of compiler optimizations is not unique to the compiler optimization field.
Phase inter-dependencies have been noted in traditional optimizing compilers between flow analysis and constant folding as well as between code generation and register allocation~\cite{leverett1979overview,Vegdahl1982}. 
In optimization theory, ``a feasible set, search space, or solution space is the set of all possible points (sets of values of the choice variables) of an optimization problem that satisfy the problem's constraints, potentially including inequalities, equalities, and integer constraints'' ~\cite{OptimizationSteuer1986multiple}.  
The optimization process normally starts by finding the initial set, and the candidates are usually pruned and down-sampled (depending on the algorithm/scenario).
Compiler optimization problems can be split into one of two subareas based on whether we (I) enable/disable a set of optimizations (optimization selection problem), or, (ii) change the ordering of those optimizations (phase-ordering problem).
Here, we briefly discuss the different optimization space of the two.
Table~\ref{tab:problemType} classifies the existing literature based on the type of the compiler optimization problem.
More recent work has addressed the selection problem since the phase-ordering problem is a more difficult research problem to solve.

\begin{table}[t!]
    \centering
        \caption{A Classification Based on the type of the Problem}
    \begin{tabular}{|>{\footnotesize\centering\arraybackslash}m{1.3cm}|>{\footnotesize\centering\arraybackslash}m{12cm}|}
        \hline
        Classes & \footnotesize{References} \\ \hline
        The Selection Problem &
  \cite{Ashouri2014bayesian,Ashouri2013VLIW,Monsifrot2002,Koseki1997,Fursin2005,Cooper2005,Cooper2002,Yuki,Park2011,Ashouri2016Res4ant,Lokuciejewski2009,Hoste2010,Kulkarni2013,Leather2009,Fraser1999,Cavazos2006,Park2015,Cavazos2005,Lokuciejewski2009,Stephenson2006,Fursin2007a,Martins2016TACO,Ashouri2016Cobayn,Hoste2008,fursin2009collective,Fursin2010,Pouchet2010,Knijnenburg2003,ashouri2012masterThesis,Fursin2002,Cheniterativecompilationdset,Pan2006,Dubach2007,Li2014,Almagor2004,Luo2014,Stephenson2003a,Park2014,Cavazos2006a,Cavazos2004,Fursin2004,Pouchet2007,Pouchet2008,Wang2009,Stephenson2003,Wolczko2000,Cavazos2006b,Vaswani2007,Fursin2007,Fursin2008,Fursin2011,Lokuciejewski2010,Ansel2014,Sarkar2000,Cooper1999,Haneda2005,Killian2014,Dubach2009,Namolaru2010,Fang2015,Stephenson2005,Franke2005,Cavazos2007,Pan2004,Thomson2009,Mars2009,Pinkers2004,Schkufza2014,Tournavitis2009,Park2012,Agakov2006,Stock2012,Sanchez2011,Zhao2003,Childers2005,bodin1998iterative,csci1800,sarkar1997automatic,tartara2013continuous,pallister2013identifying,tartara2012parallel,kelefouras2017methodology,blackmore2017automatically,Liu2018,cummins2015autotuning,ashouri2016phd,ashouri2018automatic,Ashouri2018book-dse,Ashouri2018book-cobayn}
        \\ \hline
         The Phase-ordering Problem  & 
    \cite{Ashouri2017micomp,Nobrea,Ashouri2016Res4ant,NobreRicardoLusReis2016,Martins2014,Kulkarni2004,Purini2013,Kulkarni2012,Queva2007,Park2013,Ashouri2016predictiveModeling,Vegdahl1982,Whitfield1990,Whitfield1997,Kulkarni2007,Triantafyllis2003,nobre2015use,jantz2013exploiting,kulkarni2010improving,kumar2014compiler,kulkarni2006exhaustive,nobrePhase2018,georgiou2018less,cooper2002compilation,Ashouri2018book-phase,Ashouri2018book-micomp,ashouri2016phd}
        \\ \hline
     \end{tabular}
    \label{tab:problemType}
\end{table}

\subsubsection{The Problem of Selecting the Best Compiler Optimizations}
\label{sec:introduction:optimizationSpace:selection}

Several compiler optimization passes form an optimization sequence.
When we ignore the ordering of these optimizations and focus on whether to apply the optimizations, we define the problem scope to be the selection of the best compiler optimizations.
\minor{Previous researchers have shown that the interdependencies and the interaction between enabling or disabling optimizations in a sequence can dramatically alter the performance of a running code even by ignoring the order of phases~\cite{bodin1998iterative,Agakov2006,Ashouri2016Cobayn}.}

\paragraph{Optimizations Space}
\label{sec:introduction:optimizationSpace:selection:optSpace}

Let $\optvector$ be a Boolean vector having elements $\opt$ as different compiler optimizations.
An optimization element $\opt$ can be defined as either $\opt=1$ (enabled), or $\opt=0$ (disabled).
An optimization sequence can be specified by a vector $\optvector$.
This vector is Boolean and has $\totalopt$ dimension:

\vspace{-2ex}
\begin{equation}
\label{eq:selection_space_ordinary}
\small
|\Omega_{Selection}|=\{0 , 1\}^{\totalopt}
\end{equation}  

\minor{For application $\app{_{i}}$ being optimized, $\totalopt$ shows the number of optimizations under analysis \cite{Ashouri2014bayesian}.
The selection problem's optimization space has an exponential space as its upper-bound.
For example, with $\totalopt$ = 10, there is a total state space of $2^n$ (1024) optimization options to select among.
The different optimizations scale by the total number of target applications $\app{_{i}}$, yielding a search space of $\appset = \app{_{0}} ... \app{_{N}}$ where $\appset$ is the set of  our applications under analysis.
One can define an extended version of the optimization space by switching a binary choice (on or off) to a many-choice variant per each compiler optimization. 
Equation~\ref{eq:selection_space_ordinary} shows the case where we have more than just a binary choice.
Certain compiler optimizations such as loop unrolling and tiling offer multiple constant factors of tuning, i.e., ${4, 8, 16, m}$.
Also, some optimizations can take more than one parameter.
To simplify the presentation of the equation, we consider $m$ as their total number of optimization choices a compiler optimization have.
\minor{We restrict this proposed presentation to discretized values. Continuous ranges could be approximated by predefining a discrete number of options within the continuous range of values; however, we did not find continuous options with any compilers used in this research domain.}
Consequently, we have the previous equation as:}

\vspace{-2ex}
\begin{equation}
\label{eq:selection_space_extended}
\small
|\Omega_{Selection\_Extended}|=\{0 , 1, ..., m\}^{\totalopt}
\end{equation}

\subsubsection{The Phase-ordering Problem}
\label{sec:introduction:optimizationSpace:phaseOrdering}

\minor{A problem common to multi-phase optimizing compilers is that there is no ideal ordering of phases.
An optimization pass $A$, transforms the program in ways that hinders the effect of some optimizations that otherwise could have been performed by the following pass $B$.
If the order of the two phases is switched, phase $B$ performs optimizations that may deprive phase $A$ of opportunities.
The dual of this situation is one in which the two phases open up new opportunities for each other \cite{leverett1979overview}. 
Compiler designers must take into account the order in which each optimization phase is placed and performed.
``A pair of optimization phases (comprises of many passes) may be interdependent in the sense that each phase could benefit from transformation produced by the other'' \cite{leverett1979overview,Vegdahl1982}. }

\paragraph{Optimizations Space}
\label{sec:introduction:optimizationSpace:phaseOrdering:optSpace}

A phase-ordering optimization sequence is in the factorial space due to having permutations:

\vspace{-2ex}
\begin{equation}
\label{eq:phase-ordering_space}
\small
|\Omega_{Phases}|={\totalopt}!   
\end{equation}

\noindent where $\totalopt$ shows the number of optimizations under analysis \cite{Ashouri2014bayesian,Ashouri2016predictiveModeling}.
However, the mentioned bound is a simplified phase-ordering problem having a fixed length optimization sequence length and no repetitive application of optimizations.
Allowing optimizations to be repeatedly applied and a variable length sequence of optimizations will expand the problem space to:

\vspace{-2ex}
\begin{equation}
\small
\label{eq:phase-ordering_Extendedspace}
|\Omega_{Phases\_Repetition\_variableLength}|= \sum\limits_{i=0}^l\totalopt^i
\end{equation}  

\noindent \minor{where $\totalopt$ is the number of optimizations under study and $l$ is the maximum desired length for the optimization sequence.
Even having reasonable values for $\totalopt$ and $m$, the formed optimization search space is enormous.
For example, assuming  $\totalopt$ and $m$ are both equal to 10, leads to an optimization search space of more than 11 billion different optimization sequences to identify given each piece of code being optimized.
The problem of finding the right phases does not have a deterministic upper-bound given an unbounded optimization length ~\cite{leverett1979overview,Ashouri2016predictiveModeling}.}

\subsection{Autotuning Techniques Overview}
\label{sec:introduction:optimizationSpace:autotuningTechniques}

\begin{figure*}
\centering
\includegraphics[width=\textwidth]{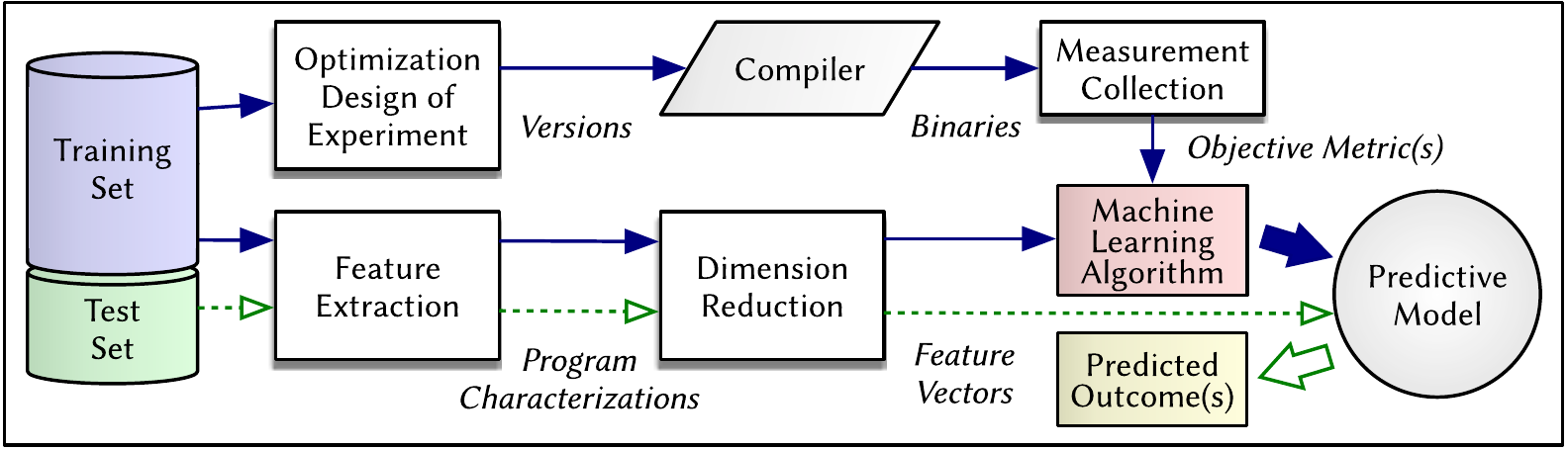}
\caption{A Sample Autotuning Framework Using Machine Learning (1) Top: The data flow through the various components of the training phase, where a model is induced based on the characteristics of applications under training set, and, (2) Bottom: The data flow through the various components of the test phase, where the already trained model is used to predict an outcome for a given test application with applications under the test set.} 
\label{fig:sampleAutotuningFramework}
\end{figure*}

\major{Autotuning refers to a methodology where there is some model \minor{(can be a search heuristic, an algorithm, etc.)} that infers one or more objectives with minimal or no interaction from a user~\cite{wilson1994suif}. 
Figure~\ref{fig:sampleAutotuningFramework} shows a sample autotuning framework leveraging compilers and machine learning.
This section will go over each of the components for this example framework.
\minor{In the domain of compilers, autotuning usually refers to defining an optimization strategy by means of a design of experiment (DoE) \cite{roy2001design} where a tuning parameter space is defined.
This parameter space creates many versions of a given input program.
When many versions of a given input program are explored, this is known as iterative exploration.
Iterative exploration can either be done exhaustively by examining all versions of a given input or sampling a subset of the space by using a  search exploration heuristic.}
When a search exploration heuristic is used, the entire search space may not be examined. This causes a trade-off between the number of instances used to construct a model (accuracy) and the computational complexity required to construct a model (time).}

\major{An autotuning framework must also have a desired target or outcome which is usually defined in terms of performance.
Extracted information from an application and measured objectives are fed into a machine learning algorithm from which a predictive model is emitted.
\minor{The extracted application information could be extremely large or may be of varying sizes.
Some form of dimension reduction is usually applied to the extracted features to reduce the amount of information. Dimension reduction is also necessary to form consistent features across varying applications}.
Principle component analysis is one of the more popular dimension reduction methods, as indicated by the surveyed research in this domain \cite{ashouri2016phd,ashouri2018automatic,Ashouri2018book-background}.}

\major{Finally, once the model is constructed, the test set can be fed through feature extraction and dimension reduction to produce a feature vector which can be fed into the predictive model.
A predicted outcome is yielded from this model and can be compared to the known outcome for evaluations and prediction error measurement.}

\section{Application characterization techniques}
\label{sec:characterization techniques}

For computer architects and compiler designers, understanding the characteristics of applications running on current and future computer systems is of utmost importance during design.
\major{Applications of machine learning to compiler optimizations require a representation of the code being optimized. Thus, a tool is required to scan the code being optimized to collect its representative features.}
To obtain a more accurate model, compiler researchers have been trying to better understand the behavior of running applications and build a vector of features that best represents pair functionality.
In general, (1) the derived feature vector must be representative enough of its application, and, (2) different applications must not have the same feature vector.
Thus, constructing a large, inefficient feature vector slows down -- or even halts -- the ML process and can reduce the precision.
In this survey, we present different program characterization techniques used in the referenced literature on compiler autotuning: (i) Static analysis of the features, (ii) Dynamic feature collection, and, (iii) Hybrid feature collection, which uses a combination of the previous two or other extraction methods.
Table~\ref{tab:app_characterization_technique} refers to such classification. 

\begin{table}[t!]
    \centering
        \caption{A Classification Based on Application Characterization Techniques}
    \begin{tabular}{|l|>{\footnotesize\centering\arraybackslash}m{2cm}|>{\footnotesize	\centering\arraybackslash}m{10cm}|}
        \hline
        \multicolumn{2}{|c|}{Classification} & {References} \\ \hline
        \multirow{2}{*}[-1ex]{\begin{turn}{90} \footnotesize{Static} \end{turn}} & 
        Source-code Features & 
   ~\cite{Monsifrot2002,Yuki2012,Yuki,Ashouri2016Res4ant,Kulkarni2013,Leather2009,Fraser1999,Cavazos2006,Park2015,Lokuciejewski2009,Stephenson2006,Martins2016TACO,Ashouri2016Cobayn,fursin2009collective,Fursin2010,Hall2009,Dubach2007,Li2014,Luo2014,Park2014,Cavazos2004,Fursin2004,Stephenson2003,Cavazos2006b,Fursin2008,Fursin2011,Kulkarni2012,Queva2007,Namolaru2010,Pan2004,Mars2009,Tournavitis2009,Park2012,Agakov2006,Sanchez2011,coons2008feature,cummins2015autotuning,Liu2018,gongempirical}
        \\
        \cline{2-3}
        & Graph-based (IR) Features  & 
   ~\cite{Whitfield1990,Nobrea,Koseki1997,Cooper2005,Leather2009,Fraser1999,Park2015,Stephenson2006,Dubach2007,Li2014,Luo2014,Park2014,Cavazos2006a,Cavazos2004,Fursin2004,Namolaru2010,Stephenson2005,Pan2004,Mars2009,Tournavitis2009,Stock2012,cummins2015autotuning}
        \\ \hline
        \multirow{2}{*}[0ex]{\begin{turn}{90} \footnotesize{Dynamic} \end{turn}} 
        & Architecture Dependent & 
   ~\cite{Fursin2005,Park2011,Park2015,pallister2013identifying,Stephenson2006,Ding2015,Cheniterativecompilationdset,Park2014,Fursin2004,Wang2009,Vaswani2007,Fursin2007,Killian2014,Dubach2009,Stephenson2005,Park2013,Cavazos2007,Pan2004,Mars2009,Tournavitis2009,Sanchez2011,Liu2018}
        \\  \cline{2-3}
        & Architecture Independent &  
   ~\cite{Ashouri2017micomp,Ashouri2014bayesian,Ashouri2016Res4ant,Ashouri2016Cobayn,Cheniterativecompilationdset,Dubach2007,Park2013,Ashouri2016predictiveModeling}
        \\ \hline
        \multicolumn{2}{|l|}{\footnotesize{Hybrid features}} & 
   ~\cite{Ashouri2016Res4ant,Leather2009,Park2015,Ding2015,Ashouri2016Cobayn,Li2014,Park2014,Pan2004,Mars2009,Tournavitis2009,Sanchez2011}
        \\ \hline
    \end{tabular}
    \label{tab:app_characterization_technique}
\end{table}

\subsection{Static Analysis}
\label{sec:characterization techniques:static}

Static Analysis, or static features collection, tries to collect features that are non-functional to a code being run on a given architecture.
Static analysis involves parsing source code at the front-end, intermediate representation (IR), the backend, or any combination of the three.
Collecting static features doesn't require the code to be executed and is considered to be one of the strongest support cases for its use.
We briefly classify source code features leveraged in prior research.

\subsubsection{Source Code Features}
\label{sec:characterization techniques:static:src}

Source code (SRC) features are abstractions of some selected properties of an input application or the current compiler intermediate state when other optimizations have already been applied.
These features range from simple information such as the name of the current function to the values of compiler parameters to the pass ordering in the current run of the compiler.
There are numerous source-code feature extractors used in the literature.
Fursin et. al. proposed Milepost GCC~\cite{Fursin2008,Fursin2011}, as a plugin to GCC compiler to extract source-level features \footnote{http://ctuning.org/wiki/index.php/CTools:MilepostGCC:StaticFeatures:MILEPOST\_V2.1}.

\subsubsection{Graph-based Features}
\label{sec:characterization techniques:static:graph}

Graph-based representation makes data and control dependencies explicit for each operation in a program.
``Data dependence graphs have provided an explicit representation of the definition-use relationships implicitly  present in a source-code'' ~\cite{ottenstein1978data,Ferrante_1987}.
A control flow graph (CFG)~\cite{allen1970control} has often been the representation of the control flow relationships of an application.
The program dependence graph explicitly represents both the essential control relationships and the essential data relationships (as present in the data dependence graph).
This information is available without the unnecessary sequencing present in the control flow graph~\cite{Ferrante_1987}.
There are numerous tools to extract a control flow graph of a kernel, a function, or an application.
MinIR~\cite{MinIR}, LLVM's Opt 
, IDA pro~\cite{eagle2011ida} are such examples of these tools available. 

Koseki et. al.~\cite{Koseki1997} used CFG and dependency graphs to understand good unrolling factors to apply to loops.
According to the formula for determining the efficiency of loop execution $P$ increases monotonically until it saturates.
Therefore, their algorithm never chooses directions that lead to local maximum points.
Moreover, they showed the method can find the point for which unrolling is performed the fewest times, as it chooses the nearest saturated point to determine the number of times and the directions in which loop unrolling is performed.

Park et. al.~\cite{Park2012} introduced a novel compiler autotuning framework which used graph-based features of an application from its intermediate representation (IR).
The authors used MinIR on the SSA form of the application to collect control flow graph of the basic blocks including the directed edges to which they have predecessor and successors. In order to construct the feature vectors needed in the predictive modeling, the authors used shortest path graph kernels ~\cite{borgwardt2005shortest}.
The method compared the similarity of two application's shortest path at each basic block.
Once the feature vectors were built, a machine learning model was created to predict the speedup of a given sequence.
Finally, the authors experimentally evaluated their proposed approach using polyhedral optimization and PolyBench suite against using source-code features and dynamic features with multiple compiler backends.


\subsection{Dynamic Characterization}
\label{sec:characterization techniques:dynamic}

Dynamic characterization involves extracting the performance counters (PC) that are used to provide information as to how well an application executes.
The data can help determine system bottlenecks and provides hints on the fine-tuning of an application's performance.
Applications can also use performance counter data to determine how many system resources to utilize.
For example, an application that is data cache intensive can be tuned by exploiting more cache locality.
Here, we briefly describe and classify the different types of collecting performance counters.
Refer to Table~\ref{tab:app_characterization_technique} for full classification on different application characterization techniques.

\subsubsection{Architecture Dependent Characterization}
\label{sec:characterization techniques:dynamic:dependent}

Recent processors provide a special group of registers that make it possible to collect several measurable events corresponding to their performance.
These events can be measured with minimal disruption to a running application and can describe several characteristics such as cache behaviors (hits and misses) and memory footprints \cite{mahlke1992effective}.
For example, in the work of Cavazos~et.~al~\cite{Cavazos2007}, there were 4 registers on the AMD Athlon that could collect performance counter events; however, 60 different events could be collected.
``By multiplexing the use of special registers, it is possible to collect anywhere between 4 and 60 types of events per run''~\cite{Cavazos2007}.
A downside of using performance counters to characterize applications is its limited reuse for other platforms.
The metrics collected are solely intended for the platform the data was collected on, reducing the platform portability.

There are tools publicly available to collect such metrics as suggested by \cite{Park2011,Fang2015,Mars2009}, such as PAPI~\cite{papiPC1999}.
PAPI is able to specify a standard application programming interface (API) on different architectures for collecting ``portable'' hardware-specific performance counters.

Monitoring these events helps the tuning process by constructing a correlation between the type of object code and the performance of that code on the target architecture.
Triantafyllis et al. \cite{Triantafyllis2003} used PFMon~\cite{jarp2002methodology}, another notable performance monitoring tool when executing programs for the Intel Itanium platform.

Cavazos et. al.~\cite{Cavazos2007} proposed an offline machine learning based model which can be used to predict a good set of compiler optimization sequences.
Their method uses performance counters to construct an application's feature vector and use them to predict good compiler optimization sequences.
The authors showed that using this method, they could outperform existing static techniques because they were able to capture memory and cache behaviors of applications. 

\subsubsection{Architecture Independent Characterization}
\label{sec:characterization techniques:dynamic:independent}

The information collected from a dynamic characterization, referred to as feature vector, is a compact summary of an application's dynamic behavior at run-time.
This information summarizes essential aspects of an application's running performance, i.e. number of cache misses or the utilization of floating point unit \cite{Cavazos2007}.
However, the information can be inaccurate or misleading if the application is run on different target architectures. Variances may exist between different targets such as cache size, execution ports, and scheduling algorithms. Architecture dependent counters, while accurate, are unable to be used in a cross-platform manner.
Because of this limitation, researchers proposed a different way of collecting dynamic behaviors which can be ported to other platform if they have the same instruction set architecture (ISA), i.e. X86\_64. This way of collecting features are known as instrumentation and are done using the dynamic program analysis tools.

Intel \emph{Pin} ~\cite{pin} is a noteworthy framework that enables an instrumented collection. \emph{Pin} is a dynamic binary instrumentation tool that analyzes a binary as it executes on a given architecture.
Microarchitecture-independent workload Characterization of Applications (MICA)~\cite{hoste2007microarchitecture} is an example of a \emph{Pintool} which is capable of collecting a number of program features at run-time. 
These collected features are independent from the microarchitecture and consider components such as the branch predictor, memory footprint, and other components that exist across various architectures instead of the specific program counters previously mentioned.

\subsection{Hybrid Characterization}
\label{sec:characterization techniques:hybrid}

The hybrid characterization technique consists of the combination of the previously known technique in a way that adds more information on the application under analysis.
In some cases~\cite{Park2014,Ashouri2016Cobayn}, hybrid feature selection can more accurately capture application behaviors since different levels of feature extraction are considered. 

Park et. al.~\cite{Park2014} used HERCULES, a pattern-driven tool to characterize an application ~\cite{kartsaklis2012hercules,kartsaklis2014hslot}.
This characterization technique is suitable for identifying interesting patterns within an application. They used HERCULES to construct prediction models that effectively select good sets of compiler optimization sequences given an application.

\subsection{Dimension Reduction Techniques}
\label{sec:characterization techniques:dimensionReduction}

\major{In the studied literature, a dimension-reduction process is important for three main reasons: (i) it eliminates the noise that may perturb further analyses, (ii) reduces the size and training time of a model, and, (iii) improved understanding of the feature space.
The techniques used are mainly Principal Component Analysis (PCA)~\cite{PCAjohnson2002}, and Exploratory Factor Analysis (EFA)~\cite{gorsuch1988FA}.} Reducing the input feature space for a machine learning algorithm may yield better learning results; however, the type of feature reduction used can be an important factor itself ~\cite{Agakov2006,Ashouri2016Cobayn}. For instance, PCA was used in the original work proposed by Ashouri et al. ~\cite{Ashouri2014bayesian}, but later the authors revised the model by exploiting EFA and observed clear benefits~\cite{Ashouri2016Cobayn}.
We elaborate more on these methods in latter sections of the survey.
Table~\ref{tab:dimension_reduction_types} classifies the use of each technique in the studied literature.


\begin{table}[t!]
    \centering
        \caption{A Classification Based on Dimension Reduction Techniques}
    \begin{tabular}{|>{\footnotesize\centering\arraybackslash}m{2.5cm}|>{\footnotesize	\centering\arraybackslash}m{10.5cm}|}
        \hline
        \footnotesize{Classification} & \footnotesize{References} \\ \hline
        Principal Component Analysis (PCA) &
   ~\cite{Ashouri2017micomp, Ashouri2014bayesian,Ashouri2013VLIW,Ashouri2016Res4ant,Park2015,Ashouri2016Cobayn,Knijnenburg2003,Cheniterativecompilationdset,Dubach2007,Park2013,Ashouri2016predictiveModeling,Thomson2009,Park2012,Agakov2006}
        \\ \hline
         Factor Analysis (FA) & 
   ~\cite{Ashouri2016Cobayn}
        \\ \hline
    \end{tabular}
    \label{tab:dimension_reduction_types}
\end{table}

Let $\acvector$ be a characterization vector having all feature data of an application's execution.
This vector represents $\acnumber$ variables to account for either the static, dynamic or a hybrid analyses. 
Let us consider a set of feature vectors representing the application characterization profiles $\appset$ consisting of $\appsetSize$ vectors $\acvector$. 
The feature vectors can be organized in a matrix $\profileMatrix$ with $\appsetSize$ rows and $\acnumber$ columns.
Each vector $\acvector$ (a row in $\profileMatrix$) includes a large set of characteristics obtained through analysis.
Many of these application characteristics (columns of  matrix $\profileMatrix$) are correlated to each other in complex ways, while other characteristics may have no apparent correlation.
``Both PCA and FA are statistical techniques aimed at identifying a way to represent $\acvector$ with a shorter vector $\rcvector$ while minimizing the information loss.
Nevertheless, they rely on different concepts for organizing this reduction'' ~\cite{Ashouri2016Cobayn,thompson2002statistical,gorsuch1988FA}.
In both cases, output values are derived by applying the dimension reduction and no longer directly representing a certain feature. 
Contrary to PCA, where the components are given by a combination of the observed features, EFA has factors represent a hidden process behind the feature generation.

\subsubsection{Principal Component Analysis}
\label{sec:characterization techniques:dimensionReduction:PCA}

The goal is to identify a summary of $\acvector$.
A second vector, $\pcavector$, of the same length of $\acvector$ ($\acnumber$) is organized by some extracted variable change.
$\pcavector$ is calculated with a linear combination of the elements in $\acvector$.
The combination of $\acvector$'s elements yielding $\pcavector$ is decided upon the analysis of the matrix $\profileMatrix$.
All elements in $\pcavector$ are orthogonal (uncorrelated) and are sorted by their variance.
Because of these two properties, the first elements of $\pcavector$ (also named principal components) carry most of the information of $\acvector$.
The reduction can be obtained by generating a vector $\rcvector$ which keeps only the $k$ most significant principal components in $\pcavector$.
Note that principal components in $\pcavector$ are not meant to have a meaning; they are only used to summarize the vector $\acvector$ as a signature.

\subsubsection{Factor Analysis}
\label{sec:characterization techniques:dimensionReduction:FA}

It explains the correlation between the different variables in $\acvector$.
Correlated variables in $\acvector$ are likely to depend on the same hidden variable in $\rcvector$.
The relationship between $\rcvector$ and the observed variables is obtained by exploiting the maximum likely method based on the data in matrix $\profileMatrix$.
When adopting PCA, each variable in $\rcvector$ tends to be composed of all variables in $\acvector$.
Because of this characteristic, it is difficult to tell what a specific component in PCA space represents.
When adopting EFA, the components $\rcvector$ tend to depend on a smaller set of elements in $\acvector$ which correlate with each other.
$\rcvector$ is a compressed representation of $\acvector$, where elements in $\acvector$ that are correlated (i.e. that carry the same information) are compressed into a reduced number of elements in $\rcvector$.

\section{Machine Learning Models}
\label{sec:ML_Models}

Machine learning investigates the study and construction of techniques and algorithms that are able to learn certain attributes from data and make predictions on them ~\cite{mohri2012MachineLearning}.
Many types and subfields of machine learning exist, so we chose to classify them based on three broad categories: (i) Supervised learning, (ii) Unsupervised learning, and, (iii) Other Methods (including reinforcement learning, graph-based technique, and statistical methods).
The classification is depicted in Table~\ref{tab:Machine_Learning_models}.
In each subsection, we provide an overview of the method and provide a summary of the notable related work.
When adapting an application of machine learning to compilers, the general optimization problem is to construct a function that ``takes as input a tuple $(F,T)$, where $F$ is the feature vector of the collected characteristics of an application being optimized'' \cite{Park2013}; and $T$ is one of the several possible compiler optimization sequences applied to this application.
Its output may be a predicted speedup value of $T$, or depending on the scenario, the predicted optimization sequence of $T$, when applied to the original code \cite{Park2013}.
There are other prediction types which we discuss and classify in details in Section \ref{sec:predication_classes}.

\begin{table}[t!]
    \centering
    \caption{A Classification Based on Machine Learning Models}
    \hspace*{-1ex}
    \begin{tabular}{@{}|>{\footnotesize\centering\arraybackslash}m{0.4cm}|>{\footnotesize\centering\arraybackslash}m{2cm}|>{\footnotesize\centering\arraybackslash}m{10.2cm}|@{}}
        \hline
        \multicolumn{2}{|c|}{Classification} & \normalsize{References} \\ \hline
        \settowidth\rotheadsize{Supervised Learning}
        \multirow{5}{*}[-2ex]{{\begin{turn}{90}\footnotesize{Supervised Learning}\end{turn}}}   
        & Bayesian Net &
   ~\cite{Ashouri2014bayesian,Ashouri2016Cobayn}
        \\ \cline{2-3}
        & Linear Models / SVMs &
   ~\cite{Ashouri2017micomp, Ashouri2016Cobayn,Ashouri2016predictiveModeling,Park2011,Park2013,Sanchez2011,Stephenson2005,Cosenza2017}
        \\ \cline{2-3}
        & Decision Trees / Random Forests &
   ~\cite{Luo2014,Fursin2007a,Ding2015,Fraser1999,Fursin2007a,Fursin2008,Fursin2011,Kulkarni2013,Monsifrot2002,Lokuciejewski2009,Park2014,Vaswani2007,bodin2016integrating}
        \\ \cline{2-3}
         & Graph Kernels  & 
   ~\cite{Park2012,Martins2016TACO,Martins2014,kindestam2017graph}
        \\  \cline{2-3}
        & Others   &
   ~\cite{Ashouri2017micomp, Ashouri2016Cobayn,Fursin2005,Park2011,Ashouri2016Res4ant,Leather2009,Park2015,Stephenson2006,Cavazos2004,Fursin2004,Wang2009,Stephenson2003,Vaswani2007,Killian2014,Namolaru2010,Park2013,Thomson2009,Tournavitis2009,Cooper2002,Ashouri2016Res4ant,Lokuciejewski2009,Kulkarni2013,Fraser1999,Cavazos2006,Stephenson2006,Ashouri2016Cobayn,fursin2009collective,Fursin2010,ashouri2012masterThesis,Pan2006,Dubach2007,Cavazos2004,Fursin2004,Wang2009,Stephenson2003,Vaswani2007,Fursin2007,Fursin2008,Fursin2011,Kulkarni2012,Lokuciejewski2010,Ansel2014,Haneda2005,Dubach2009,Namolaru2010,Fang2015,Franke2005,Cavazos2007,Pinkers2004,Schkufza2014,Agakov2006,Stock2012,Sanchez2011,csci1800,Cooper1999,cummins2015autotuning}
        \\ \cline{2-3}
        \hline
        \settowidth\rotheadsize{UnS}
        \multirow{3}{*}[0.5ex]{{\begin{turn}{90} \scriptsize{Unsupervised}  \end{turn}}}
        & Clustering  &
        
   ~\cite{Ashouri2017micomp,Ashouri2013VLIW,Thomson2009,Purini2013,Martins2014,Martins2016TACO}

        \\ \cline{2-3}
         & Evolutionary Algorithms (GAs,NEAT,NN) &
   ~\cite{Cooper1999,Cooper2005,Cooper2002,Lokuciejewski2009,Hoste2010,Kulkarni2013,Leather2009,Cavazos2006,Park2015,Cavazos2005,Stephenson2006,Hoste2008,Knijnenburg2003,NobreRicardoLusReis2016,Martins2014,Dubach2007,Kulkarni2004,Li2014,Almagor2004,Purini2013,Stephenson2003a,Pouchet2008,Wang2009,Stephenson2003,Kulkarni2012,Park2012,Agakov2006,Kulkarni2007,Garciarena:2016:EOC:2908961.2931696,kulkarni2010improving,Nobrea,Stephenson2006,coons2008feature,Kulkarni2012,Kulkarni2013,tartara2013continuous,Garciarena:2016:EOC:2908961.2931696,falch2015machine,kumar2014compiler,cummins2017deep,gccAutomatic2018}
        
        \\ \cline{2-3}
               \hline
        \multicolumn{2}{|l|}{\footnotesize{Reinforcement Learning}} &
    \cite{mcgovern2002building,mcgovern1999scheduling,coons2008feature}
        \\ \hline
    \end{tabular}
    \label{tab:Machine_Learning_models}
\end{table}

\subsection{Supervised learning}
\label{sec:ML_Models:supervised}

Supervised learning is the subclass of machine learning that deals with learning a function from labeled data in the training set ~\cite{dietterich2000ensemble,mohri2012MachineLearning}.
The learner receives a set of labeled examples as training data and makes predictions for all unseen points.
This is the most common scenario associated with \emph{classification}, \emph{regression}, and \emph{ranking} problems.

\subsubsection{Bayesian Networks}
\label{sec:ML_Models:supervised:BN}

Bayesian Networks (BN) ~\cite{pearl1985bayesian,friedman1997bayesian} are powerful classifiers to infer the probability distribution variables, which can be binary as well, that characterize a certain attribute such as the optimality of compiler optimization sequences. 
``A Bayesian Network is a direct acyclic graph whose nodes are variables and whose edges are the inter-dependencies between them''~\cite{Ashouri2014bayesian,Ashouri2016Cobayn}.
Computed probabilities of those observed variables categorized under nodes can be further used as input to the model as \emph{evidence}.
\major{The probability distributions of compiler optimizations depend on the program features and the relationship between the different optimizations that are applied.}
Ashouri et. al. ~\cite{Ashouri2014bayesian,Ashouri2016Cobayn} proposed a Bayesian Network approach to infer the best compiler optimizations suitable for an embedded processor. The approach uses static, dynamic, and hybrid features of an application as evidence to the trained networks. Rather using all training data, the authors proposed to base their training data on only those instances of compiler sequences having the best 15\% speedup w.r.t the GCC's standard optimization level \texttt{-O3}. This way, BN is trained with only good sequences of optimizations and subsequently ables to populates the optimal solutions for an unseen application. The evaluations were carried out with Cbench~\cite{fursin2010collective} and Polybench ~\cite{polyOrig,polyJohn} suites and the authors showed that employing BN with an iterative compilation outperforms GCC's \texttt{-O2} and \texttt{-O3} by around 50\%.

\subsubsection{Linear Models and SVMs}
\label{sec:ML_Models:supervised:linearAndSVMs}

Linear models are one of the most popular supervised learning methods to be widely used by researchers in tackling many machine learning applications.
``Linear regression, nearest neighbor, and linear threshold algorithms are generally very stable'' ~\cite{dietterich2000ensemble}.
Linear models are algorithms whose output classifier does not yield dramatic fluctuation against minor changes in the training set. 
Moreover, ``SVMs are a type of supervised machine learning technique, can be used for both regression and classification scenarios'' \cite{Park2011}. 
In SVMs, there are algorithms that can adapt linear methods to a non-linear class of problems using a technique called kernel trick \cite{scholkopf2001kernel}. 
Aside from constructing hyperplane or a set of hyperplanes in an high-dimensional space, SVMs can find the best hyperplane (so-called maximum margin clustering ~\cite{kim2003financial,Park2013}).

Sanchez et al. \cite{Sanchez2011} proposed to SVMs to learn models to focus on autotuning the JIT compiler of IBM Testarossa and the build compilation plan.
They used scalar features to construct feature vectors and employed the learning scheme of SVMs to experimentally test the quality of their model using a single-iteration and 10-iteration scenarios on \texttt{SPECjvm98} benchmark suite.

\subsubsection{Decision Trees and Random Forests}
\label{sec:ML_Models:supervised:DSandRF}

A binary decision tree is a tree representation of a partition of the feature space.
Decision trees can be defined using more complex node questions resulting in partitions based on more complex decision surfaces~\cite{mohri2012MachineLearning}.
Random forests, or random decision forests, are an ensemble learning method that use multiple learning methods to provide a better prediction for regression and classification purposes.
Random forests start by constructing many decision trees at training time and output the class that fits the mode of classes (in the case of classification) or means of classes (in the case of regression).
``Random decision forests correct decision trees' habit of overfitting to their training set'' ~\cite{friedman2001elements,dietterich2000ensemble}.

Fraser et al.~\cite{Fraser1999} proposed to use machine learning to perform code compression.
It used the IR structure of codes to automatically infer a decision tree that separates intermediate representation code into a stream that compresses more efficiently.
They evaluated their code compression approach with GCC and used opcodes which can also help predict elements of the operand stream.

Monsifrot~\cite{Monsifrot2002} addressed the automatic generation of optimization heuristics for a target processor through machine learning. They used decision trees to learn the behavior of the loop unrolling optimization on the code being studied to decide whether to unroll on UltraSPARC and IA-64 machines.


\subsubsection{Graph Kernels}
\label{sec:ML_Models:supervised:graph}

Graph kernels construct a low-dimensional data representation by a cost function that preserves properties of the data.
``The use of kernel functions is very attractive because the input data does not always need to be expressed as feature vectors'' ~\cite{Koseki1997,Park2012,Nobrea}.
Graph kernels are emerging as a means of exploiting many different machine learning applications on a wide range of applications from semi-supervised learning~\cite{chapelle2009semiGraph} to clustering and classification~\cite{camps2007semi}.

Park et. al.~\cite{Park2012} proposed the use of graph kernels to characterize an application.
The authors used the control flow graph of programs.
Instead of flattening the control flow graph information into a fixed-length characterization vector, the authors created a similarity matrix to feed into SVM algorithm using the shortest path graph kernel ~\cite{borgwardt2005shortest}.
The output of the shortest path graph kernel is a similarity score between two graphs.
They evaluated the approach by using a model for both non-iterative and iterative fashion on two multicore systems.
GCC and Intel compiler were the two back-end compilers once the code passed through a polyhedral source-to-source compiler.
Finally, they compared the proposed model against using static features derived by MilepostGCC, performance counters, and 5-Reactions. 


\subsubsection{Other Supervised Methods}
\label{sec:ML_Models:supervised:others}

For conciseness proposes, we decided to classify other supervised learning methods under this subsection. These include Neural Networks, ANOVA, K-nearest neighbor, Gaussian process learning~\cite{mohri2012MachineLearning}, etc..

Moss et al.~\cite{cavazos1998} showed a machine learning process to construct heuristics for instruction scheduling, more specifically, scheduling a straight-line code. 
They used static and IR features of the basic block with the \texttt{SPEC} benchmark to experimentally evaluate their approach by using Geometric mean as fitness function and fold-cross-validation.

Cavazos and Moss~\cite{Cavazos2004} used the JIT Java compiler and SPECjvm98 benchmarks with a rule set induction learning model to decide whether to schedule instructions.
They exploited supervised learning to induce heuristics to predict which blocks should schedule to maximize the benefit where the induced function acts as a binary classification.
The authors experimentally showed they could obtain at least 90{\%} of scheduling improvement for every block while spending at most 25{\%} of the needed effort.


Haneda et al.~\cite{Haneda2005} introduced a statistical model to build a methodology which reduces the size of the optimization search space, thus allowing a smaller set of solutions for exploring strategies.
They showed that the technique found a single compiler sequence across a set of SPECint95 benchmarks which could perform better on average against the GCC's standard optimization set.

Tournavitis et al.~\cite{Tournavitis2009} proposed a technique using profile-driven parallelism detection in which they could improve the usability of source code features.
The proposed approach enabled the authors to identify and locate more parallelism on an application with user feedback required at the final stage.
Moreover, the authors exploit a machine-learning based prediction mechanism that replaces the target-specific mapping heuristic.
This new mapping scheme made better mapping decisions while being more scalable and practical to use on other architectures.

Namolaru et al.~\cite{Namolaru2010} proposed a general method for systematically generating numerical features from an application.
The authors implemented the approach on GCC.
This method does not place any restriction on how to logically and algebraically aggregate semantical properties into numerical features.
Therefore, it  statistically covers all relevant information that can be collected from an application.
They used static features of MilePost GCC and MiBench to evaluate their approach.

\subsection{Unsupervised learning}
\label{sec:ML_Models:Unsupervised}

Unsupervised learning is the machine learning task of inferring a function to describe hidden structure from unlabeled data.
Since the examples given to the learner are unlabeled, there is no error or reward signal to evaluate a potential solution~\cite{hastie2009unsupervised,mohri2012MachineLearning}.
Unsupervised learning is closely tightened with the problem of density estimation in statistics~\cite{silverman1986density}; however, unsupervised learning also encompasses many other techniques that seek to summarize and explain key features of the data such as the use of evolutionary algorithms.
In unsupervised learning targets don't exist, but an environment that permits model evaluation does exist. An environment can score a model with some value, i.e. which is the value passed to a model's objective function. 

\subsubsection{Clustering Methods}
\label{sec:ML_Models:Unsupervised:clustering}

One of unsupervised learning's key subclasses is clustering.
Clustering helps to downsample the chunk of unrelated compiler optimization passes into meaningful clusters that correspond to each other, i.e. targets loop-nests or scalar values, or they should follow each other in the same sequence.
The other importance of clustering and downsampling is to reduce the compiler optimization space, which can be tens of thousands orders of magnitude smaller (mentioned in Section~\ref{sec:introduction:optimizationSpace}).

Thomson et al. ~\cite{Thomson2009} proposed a methodology to decrease the training time of a machine learning based autotuning.
They proposed to use a clustering technique, namely Gustafson Kessel algorithm \cite{babuka2002improved}, after applying the dimension reduction process.
They evaluated the clustering approach on the EEMBCv2 benchmark suite and showed a factor of seven on reducing the training time with the proposed approach.

Ashouri et al. ~\cite{ashouri2012masterThesis,Ashouri2013VLIW} developed a hardware/software co-design toolchain to explore compiler design space jointly with microarchitectural properties of a VLIW processor.
The authors have used clustering to derive to four good hardware architectures followed by mitigating the selection of promising compiler optimization with statistical techniques such as Kruskal-Wallis test and Pareto-optimal filtering.
(This method involved with statistical methods as well. Refer to Section~\ref{sec:ML_Models:supervised:others})

Martins et al. ~\cite{Martins2014,Martins2016TACO} tackled the problem of phase-ordering by a clustering-based selection method for grouping functions with similarities. 
Authors used DNA encoding where program elements (e.g., operators and loops in function granularity) are encoded in a sequence of symbols and followed by calculating the distance matrix and a tree construction of the optimization set.
Finally, they applied the optimization passes already included in the optimization space to measure the exploration speedup versus the state-of-the-art techniques such as Genetic algorithm.

\subsubsection{Evolutionary Algorithms}
\label{sec:ML_Models:Unsupervised:GA}

Evolutionary algorithms are inspired by biological evolution such as the process of natural selection and mutation.
Candidate solutions in the optimization space play the role of individuals in a population.
A common practice to identify the quality of solutions is using a fitness function such as an execution speedup.
Evolution of the population takes place after the repeated application of the fitness function \cite{mohri2012MachineLearning}.
Here we briefly mention some of the more notable techniques used in the literature. 

Genetic Algorithm (GA) is a meta-heuristic algorithm under this class which can be paired with any other machine learning technique or be used independently.
\major{A notable fast GA heuristic is NSGA-II (Non-dominated Sorting Genetic Algorithm II) \cite{deb2002fast}, which is a popular method for many multi-objective optimization problems and have had numerous applications mostly in the computer architecture domain \cite{mariani2010correlation,silvano2011multicube}. NSGA-II is shown to alleviate the computational complexity of classic GA algorithms}. 

\major{Another interesting evolutionary model is Neuro Evolution of Augmenting Topologies (NEAT) ~\cite{stanley2002efficient}.
They proved to be a powerful model for learning complex problems since they are able to change the network topology and parameter weight to find the best-balanced fitness function.
NEAT specifically has been used in many notable recent research work as well \cite{coons2008feature,Kulkarni2012,Kulkarni2013}.
This section summarized a few notable research work that used evolutionary algorithms.}

Cooper et al.~\cite{Cooper1999,Cooper2002} addressed the code size issue of the generated binaries by using genetic algorithms.
The results of this approach were compared to an iterative algorithm generating fixed optimization sequence and also at random frequency.
Given the comparison, the authors concluded that by using their GAs they could develop new fixed optimization sequences that generally work well on reducing the code-size of the binary.

\major{Knijnenburg et al.\cite{Knijnenburg2003} proposed an iterative compilation approach to tackle the selection size of the tiling and the unrolling factors in an architecture independent manner. The authors evaluated their approach using a number of state-of-the-art iterative compilation techniques, e.g., simulated annealing and GAs, and a native Fortran77 or \texttt{g77} compiler enabling optimizations for Matrix-Matrix Multiplication (M$\times$M), Matrix-Vector Multiplication (M$\times$V), and Forward Discrete Cosine Transform.}



Agakov et al.~\cite{Agakov2006} adapted a number of models to speed up the exploration of an iterative compilation space. The methodology exploits a Markov chain oracle and an independent identically distributed (IID) probability distribution oracle.
The authors gained significant speedup by guiding their iterative compilation using these two models when tested on unseen applications. When predicting the best optimizations given an unseen application, they use a nearest neighbor classifier. First, it identifies the training application having the smallest Euclidean distance in the feature vector space (derived by PCA). Then, it learns the probability distribution of the best compiler optimizations for this neighboring application either by means of the Markov chain or IID model.
The probability distribution learned is then used as the predicted optimal distribution for the new application. The Markov chain oracle outperformed the IID oracle, followed by the RIC methodology using a uniform probability distribution.

\major{Leather et al.~\cite{Leather2009} used a grammatical evolution-based approach with a genetic algorithm to describe the feature space.
The authors showed that the selection of the right features for a compiler can lead to substantial performance improvement to the quality of machine learning heuristic.
They described the space formed by the features with a grammar and then they generated many features from it.
These generated features are later used in the predictive modeling to search the optimization space using GCC and Mediabench and experimentally validate the proposed approach on a Pentium machine.}

Kulkarni et al. developed two approaches in order to tackle both optimization selection~\cite{Kulkarni2013} and phase-ordering ~\cite{Kulkarni2012}.
The approach for selecting the good compiler passes is done using NEAT and static features to tune the Java Hotspot compiler with SPEC Java benchmarks (using two benchmarks for training and two for testing).
The authors used NEAT to train decision trees for the caller and the callee whether to inline.
When addressing the phase-ordering problem, they proposed an intermediate speedup prediction method that used static features of the current state of the application being studied to query the model and induce the current best optimization to use.
This way, iteratively a compiler sequence is formed on an application-based manner.

Purini et al.~\cite{Purini2013} defined a machine learning based approach to downsample a set of compiler optimization sequences within LLVM's \texttt{-O2} and applied machine learning to train a model. The authors introduced a clustering algorithm to cluster sequences based on the similarity matrix by calculating the Euclidean distance between the two sequence vectors. In the experimental evaluation, they have mentioned the most frequent optimization passes with their fitness function (execution speedup) as well. Later, Ashouri et al. \cite{Ashouri2017micomp} used such similarity matrix for LLVM's \texttt{-O3} passes to speed up the exploration, or recommendation, of predicted sequences.

\subsection{Reinforcement Learning}
\label{sec:ML_Models:RL}

Reinforcement learning (RL) is an area of machine learning which can not be classified as supervised or unsupervised.
It is inspired by behaviorist psychology and uses the notion of rewards or penalties so that a software agent interacts with an environment and maximizes his cumulative reward.
The interesting difference in RL is that the training and testing phases are intertwined ~\cite{mohri2012MachineLearning}.
RL uses Markov decision process (MDP) \cite{howard1960dynamic} to adapt and interact with its environments.
In this section, we have provided the works done with RL in the field. 

McGovern et al. \cite{mcgovern1999scheduling} presented two methods of building instruction scheduler using rollouts, an improved Monte Carlo search \cite{tesauro1996line}, and a reinforcement learning.
The authors showed that the combined reinforcement learning and rollout approach could outperform the commercial Compaq scheduler on evaluated benchmarks from SPEC95 suite. 

Coons et al. \cite{coons2008feature} used NEAT as a reinforcement learning tool for finding good instruction placements for an EDGE architecture.
The authors showed that their approach could outperform state-of-the-art methods using simulated annealing in order to find the best placement.

\section{Prediction Types}
\label{sec:predication_classes}

The major classes of prediction type are: (i) to predict the right set of compiler optimizations to be used, (ii) to predict the speedup of a compiler optimization sequence without actually executing the sequence, (iii) to predict the right set of features to be used in order to characterize the application, (iv) the intermediate speedup (tournament) prediction, and, (v) the reduction of the search space through down-sampling of optimizations.
Table~\ref{tab:prediction_type} shows the classification of the related literature.

\begin{table}[t]
    \centering
        \caption{A Classification Based on Prediction Types}
    \begin{tabular}{|>{\footnotesize\centering\arraybackslash}m{1.75cm}|>{\footnotesize\centering\arraybackslash}m{11cm}|}
        \hline
        Classification & References \\ \hline
        Clustering / Downsampling &
   ~\cite{Ashouri2017micomp, Ashouri2013VLIW,Thomson2009,Purini2013,Martins2014,Martins2016TACO,Nobrea,Hoste2008,Knijnenburg2003,nobre2015use,georgiou2018less}
        \\ \hline
         Speedup Prediction  & 
   ~\cite{Ashouri2017micomp, Fursin2005,Cooper2005,Park2011,Ashouri2016Res4ant,Kulkarni2013,Leather2009,Cavazos2006,Park2015,Stephenson2006,Ding2015,Fursin2007a,Ashouri2016Cobayn,Dubach2007,Park2014,Fursin2004,Wang2009,Stephenson2003,Killian2014,Park2013,Ashouri2016predictiveModeling}
        \\ \hline
         Compiler Sequence Prediction & 
   ~\cite{bodin1998iterative,Ashouri2014bayesian,Ashouri2013VLIW,Monsifrot2002,Cooper2005,Cooper2002,Park2011,Ashouri2016Res4ant,Hoste2010,Cavazos2006,Park2015,Cavazos2005,Lokuciejewski2009,Stephenson2006,Fursin2007a,Martins2016TACO,Ashouri2016Cobayn,Hoste2008,fursin2009collective,Fursin2010,Knijnenburg2003,NobreRicardoLusReis2016,ashouri2012masterThesis,Fursin2002,Cheniterativecompilationdset,Martins2014,Kulkarni2004,Almagor2004,Purini2013,Luo2014,Cavazos2006a,Cavazos2004,Pouchet2007,Pouchet2008,Wang2009,Stephenson2003,Cavazos2006b,Vaswani2007,Fursin2007,Fursin2008,Fursin2011,Kulkarni2012,Lokuciejewski2010,Cooper1999,Haneda2005,Queva2007,Dubach2009,kulkarni2006exhaustive,Namolaru2010,Fang2015,Stephenson2005,Franke2005,Cavazos2007,Pan2004,Thomson2009,Mars2009,Pinkers2004,Schkufza2014,Tournavitis2009,Park2012,Agakov2006,Stock2012,Sanchez2011,Kulkarni2007,kulkarni2010improving,nobre2015use,csci1800,blackmore2015logic,bodin2016integrating,Liu2018,asher2017study}
        \\  \hline
         Tournament / Intermediate  &  
   ~\cite{Triantafyllis2003,Ashouri2016predictiveModeling,Kulkarni2012,Park2011,Park2015,Park2013}
        \\ \hline
        Feature Prediction & 
   ~\cite{Park2015,Ding2015,Li2014,Vaswani2007,coons2008feature,cummins2017deep}
        \\ \hline
    \end{tabular}
    \label{tab:prediction_type}
\end{table}

\subsection{Clustering and Downsampling}
\label{sec:predication_classes:clustering}

Clustering, or a cluster analysis, is the task of investigating the similarities between a set of variables in such a way that similar variables can be placed into the same group called a cluster.
Clustering is most commonly used in the field of data mining and a common technique for statistical data analysis.
Cluster analysis is used in many fields including unsupervised machine learning, pattern recognition, and compiler autotuning~\cite{mohri2012MachineLearning}.
In order to approach reasonable solutions to the compiler optimization problems, the optimization search space needs to be reduced.
Researchers try to find ways to decrease the enormous size of the optimization space by orders of magnitudes. This technique, known as downsampling, has been used in many recent works. We mention these papers in the table \ref{tab:prediction_type}.


\major{Ashouri et al. \cite{Ashouri2017micomp} presented a full-sequence speedup predictor for the phase-ordering problem that rapidly converges to optimal points and outperforms both standard optimization levels and the state-of-the-art \emph{ranking} approach \cite{Park2012,Park2013}.
The authors applied a clustering technique on all optimization passes available in LLVM's \texttt{-O3} and trained their model using the resulted subsequences.
The infeasibly large phase-ordering space is reduced to a fairly big yet explorable using iterative compilation and a machine learning method.
The authors showed their approach could outperform LLVM's highest optimization level and other existing speedup predictors with just a few predictions.}

\subsection{Speedup Prediction}
\label{sec:predication_classes:spPrediction}

Speedup predictive modeling is the process of constructing, testing, and validating a model to predict an unobserved timing outcome.
The model is constructed based on the characterization of a state.
The state being characterized is the code being optimized and the predicted outcome corresponds to the speedup metric calculated by normalizing the execution time of the current optimization sequence by the execution time of the baseline optimization sequence.
``The general form of a speedup predictor is to construct a function that takes as input a tuple $(F,T)$, where $F$ is the collected feature vector of the of an application being optimized'' \cite{Park2013}.
$T$ can be one of the many possible compiler optimization sequences in the design space.
The model's output is the predicted value of a speedup that the sequence $T$ is to achieve when applied to the application's source-code in its original state.
This prediction type is one of the more widely used among the researchers and here we mention some of its notable usages. 

Dubach et al.~\cite{Dubach2007,Dubach2009} proposed a speedup predictor based on the source code features of the optimized applications.
The authors used static features from SUIF compiler infrastructure~\cite{wilson1994suif} for VLIW and compared the result with non-feature-based alternative predictors such as mean predictors, sequence encoding-based predictors, and reaction based predictors.

Park et al.~\cite{Park2011,Park2012,Park2013,Park2014,Park2015} proposed several predictive modeling methodologies to tackle the problem of selecting the right set of compiler optimizations.
In ~\cite{Park2012}, the authors used Control Flow Graphs (CFGs) with graph kernel learning to construct a machine learning model.
First, they constructed CFGs by using the LLVM compiler and convert the CFGs to Shortest Path Graphs (SPGs) by using the Floyd-Warshall algorithm. 
Then, they apply the shortest graph kernel method~\cite{borgwardt2005shortest} to compare each one of the possible pairs of the SPGs and calculate a similarity score of two graphs.
They calculated similarity scores for all pairs are saved into a matrix and directly fed into the selected machine-learning algorithm, specifically SVMs in their work. 
Later in ~\cite{Park2013}, instead of using hardware-dependent features, authors used static features from the source-code on a polyhedral space and predict the performance of
each high-level complex optimization sequence with trained models.
In a later work ~\cite{Park2014}, they used user-defined patterns as program features, named HERCULES~\cite{kartsaklis2012hercules}, to derive arbitrary patterns coming from users.
They focused on defining patterns related to loops: the number of loops having memory accesses, having loop-carried dependencies, or certain types of data dependencies.
These works use static program features mainly focusing on loop and instruction mixes. 

\subsection{Compiler Sequence Prediction}
\label{sec:predication_classes:flagPrediction}

A compiler sequence predictor is a type of prediction model which outputs the best set of compiler passes or sequences to apply on a given application.
Application characterization is fed to this model, and the model induces a prediction of a set of compiler passes $\optimalvector \in \optspace$ to apply to the given application.
The objective could be configurable -- it could maximize its performance or fulfill other objectives such as code size or energy consumption. ~\cite{Park2011,Ashouri2014bayesian,Ashouri2016Cobayn}. 
This specific prediction type has attracted the most interest among research and we have seen the bulk of work addressing this. Here, we discuss a few notable contributions.


Cavazos et al.~\cite{Cavazos2007} investigated the problem of selecting the right set of compiler optimizations by means of dynamic characteristics of an application under analysis. The authors used logistic regression for model construction and showed that using their approach could outperform existing techniques used by static code features and rapidly reaching the achievable speedup.

Ashouri et. al~\cite{Ashouri2014bayesian,Ashouri2016Cobayn} proposed a Bayesian Network (BN) approach which was fed by either of static, dynamic or a hybrid characterization vector of an application. The BN model could subsequently induce a probabilistic model.
The authors combined their model with iterative compilation to derive good compiler sequences and experimentally showed they could outperform the state-of-the-art models.

\subsection{Tournament and Intermediate Prediction}
\label{sec:predication_classes:tournametIntermdiatePrediction}

A tournament predictor, proposed by \cite{Park2011}, predicts a binary classification between two input optimization sequences.
This type of predictor can rank compiler sequences accordingly and select the best among them.
On the other hand, intermediate speedup predictor (also referred to as Reaction-based modeling~\cite{Cavazos2006}) \cite{Kulkarni2012,Ashouri2016predictiveModeling} is a prediction model that tends to iteratively predict the speedup of the current state of an application being optimized.
Applications' characteristics in each state along with the compiler sequence $T$ serves as input, and the model predicts the speedup $T$ should achieve.
Since intermediate speedup predictors behave more or less the same as a tournament predictor, as both work on individual optimizations at a time to be applied on the current state, we decided to group these two under one classification. 
An intermediate sequence approach needs multiple execution profiles of the application being optimized; therefore, it tends to be slower in comparison to other methods. Nevertheless, it has been shown an effective method specifically to tackle the phase-ordering problems ~\cite{Kulkarni2012,Ashouri2016predictiveModeling}.

Park et al.~\cite{Park2011} have proposed tournament predictors in order classify whether an optimization should be applied as an immediate optimization.
Park et al. evaluated three prediction models using program counters (PC) derived by PAPI ~\cite{papiPC1999}: sequence, speedup, and tournament  on several benchmarks (Polybench, NAS, etc.) using the Open64 compiler.
They showed on many occasions tournament predictors can outperform other techniques.

Kulkarni and Cavazos~\cite{Kulkarni2012} developed an intermediate speedup predictor using NEATs that selects the best order of compiler optimization passes iteratively based on its current state. They evaluated their approach using JAVA Jikes dynamic compilers and observed on average 5-10 \% speedup when adjusting the different ordering of the phases.

Ashouri et al.~\cite{Ashouri2016predictiveModeling} demonstrated a predictive methodology in order to predict the intermediate speedup obtained by an optimization from the configuration space given the current state of the application. The authors used dynamic features of an application for their prediction model unlike the work of Kulkarni and Cavazos ~\cite{Kulkarni2012}. They also used speedup values between the execution times of the program before and after the optimization process as their fitness function and defined heuristics to traverse the optimization space.

\subsection{Feature Prediction}
\label{sec:predication_classes:ftPrediction}

Constructing a model from empirical data must have minimal user intervention.
The accuracy of predictions is heavily related to the type of features and characterization collected from an application under optimization process.
Therefore, feature prediction models are advantageous to use for learning the most promising features affecting performance.
During this process, choosing the right set of features to characterize an application is crucial.

Vaswani et al. \cite{Vaswani2007} used empirical data to construct a model that is sensitive to microarchitecture parameters and compiler optimization sequences.
The proposed model automatically learns the relationship from data and constructs automatic heuristics. The authors evaluated their approach using \texttt{SPEC-CPU2000} and GCC compiler and on average attained performance improvement over 10\% higher than the highest standard optimization levels available with GCC.

Li et al.~\cite{Li2014} used a  machine learning based optimization focused on feature processing. The authors adapted this method based on an application under analysis and apart from predefined source code features. They developed an approach to automatically generate several features of an application to be fed to a model constructor using a template. 
Furthermore, they observed fluctuation on the values of the features when using different compiler optimizations and alleviated this by designing a feature extractor to iteratively extract the required features at runtime to be fed to the optimization model. 

Ding et al. ~\cite{Ding2015} proposed an autotuning framework capable of incorporating different inputs in a two-level approach to overcome the challenge of input sensitivity.
They leveraged the Petabricks language and compiler ~\cite{Ansel2009} and used an input-aware learning technique to differentiate between inputs.
The work clustered the space and chose its centroid for autotuning (i) to identify a set of configurations for each class of inputs, and, (ii) to identify a production classifier which allowed them to efficiently identify the best landmark optimization to use for a new input.

\section{Optimization Space Exploration Techniques}
\label{sec:DSE}

As previously mentioned in Section~\ref{sec:introduction:optimizationSpace}, traversing the large compiler optimization space made by the different combinations of optimizations requires a proper exploration strategy.
Iterative compilation and genetic algorithms are among the most mentioned strategies in the literature.
In a broader perspective, the strategies are derived by the desired type of optimization space exploration. 
\major{Design space exploration (DSE) is the activity of exploring design alternatives before implementation.
Historically, it has been used at system-level design, but the methodologies can be adapted to use at compiler-level exploration as well \cite{Ashouri2013VLIW}.
DSE helps to define and follow exploration policies that undertake generation of candidates within certain or all optimization space and has been proven useful for many optimization tasks \cite{ku1992design} }.
In general, different applications could impose different energy and performance requirements.
The main goal of this phase is to efficiently traverse and configure the exploration parameters~\cite{palermo2003dse,palermo2005multi}.
In this section, we classify the different exploration strategies used by researchers in literature to overcome this challenge.
Table~\ref{tab:spaceExplotation_type} represents our fine-grain classification of the different related works based the exploration type.

\begin{table}[t]
    \centering
        \caption{A Classification Based on Space Exploration Methods}
    \begin{tabular}{|>{\footnotesize\centering\arraybackslash}m{1.5cm}|>{\footnotesize\centering\arraybackslash}m{11.5cm}|}
        \hline
        Classification & References \\ \hline
        Adaptive Compilation &   ~\cite{Cooper2005,Cooper2002,Childers2005,Lokuciejewski2009,Hoste2010,Leather2009,Stephenson2006,Ding2015,Fursin2007a,Hoste2008,fursin2009collective,Fursin2010,Pan2006,Kulkarni2004,Almagor2004,Luo2014,Fursin2004,Stephenson2003,Wolczko2000,Cavazos2006b,Vaswani2007,Fursin2007,Fursin2008,Fursin2011,Kulkarni2012,Lokuciejewski2010,Ansel2014,Sarkar2000,Cooper1999,Haneda2005,Ansel2009,Dubach2009,Namolaru2010,Fang2015,Stephenson2005,Franke2005,Cavazos2007,Pan2004,Mars2009,Pinkers2004,Stock2012,Sanchez2011,waterman2006adaptive,ogilvie2017minimizing,bodin2016integrating}
        \\ \hline
         Iterative Compilation  & 
   ~\cite{aarts1997oceans,bodin1998iterative,Ashouri2017micomp, Ashouri2014bayesian,Ashouri2013VLIW,Nobrea,Koseki1997,Kulkarni2007,Fursin2005,Cooper2005,Cooper2002,Yuki2012,Yuki,Park2011,ashouri2012masterThesis,Ashouri2016Res4ant,Lokuciejewski2009,Hoste2010,Kulkarni2013,Leather2009,Cavazos2006,Park2015,Cavazos2005,Stephenson2006,Ding2015,Fursin2007a,Martins2016TACO,Ashouri2016Cobayn,Hoste2008,fursin2009collective,Fursin2010,Pouchet2010,Knijnenburg2003,Chen2005,NobreRicardoLusReis2016,ashouri2012masterThesis,Fursin2002,Cheniterativecompilationdset,Martins2014,Pan2006,Dubach2007,Kulkarni2004,Almagor2004,Luo2014,Park2014,Cavazos2006a,Fursin2004,Pouchet2007,Pouchet2008,Wang2009,Stephenson2003,Wolczko2000,Cavazos2006b,Vaswani2007,Fursin2007,Fursin2008,Fursin2011,Kulkarni2012,Lokuciejewski2010,Ansel2014,Sarkar2000,Cooper1999,Haneda2005,Killian2014,Ansel2009,Dubach2009,Namolaru2010,Fang2015,Stephenson2005,Park2013,Ashouri2016predictiveModeling,Franke2005,Cavazos2007,Pan2004,Thomson2009,Mars2009,Pinkers2004,Schkufza2014,Tournavitis2009,Park2012,Agakov2006,Sanchez2011,Lokuciejewski2010,nobre2015use,kulkarni2009phaseOrdering,kulkarni2010improving,blackmore2015logic,kulkarni2006exhaustive,tartara2012parallel,tartara2013,kumar2014compiler,ogilvie2017minimizing,kelefouras2017methodology,blackmore2017automatically,Liu2018}
        \\ \hline
         Non-iterative  & 
   ~\cite{Vegdahl1982,Whitfield1990,mahlke1992effective,Whitfield1997,Fraser1999,Stephenson2006,Pouchet2010,Pouchet2008,Queva2007}
        \\  \hline
    \end{tabular}
    \label{tab:spaceExplotation_type}
\end{table}

\subsection{Adaptive Compilation}
\label{sec:DSE:adaptive}

Adaptive optimization \cite{waterman2006adaptive}, also known as profile-guided optimization, is a technique where the optimization space is explored based on the outcome of fitness functions, e.g., execution time to profile the executable and dynamically modifies/recompiles certain segments of an application under optimization.
The profiling provides enough features so the compiler can decide on what portion of the code to be recompiled.
An adaptive compiler is placed between a just-in-time (JIT) compiler and an interpreter of instructions. 
As suggested by \cite{Cooper2005}, an adaptive compiler can benefit from a compilation-execution feedback loop to select the best optimization that satisfies the objectives of a scenario.
The following works go over the state of the art and practice with profile-guided optimization.

Cooper et al.~\cite{Cooper2002,Cooper2005} developed an adaptive compiler, ACME, and a user-interface to control the process of recompilation and exploration of different compiler optimizations.
The authors introduced virtual execution -- a technique to mitigate the concurrent execution of code being optimized.
By running an application once, this technique helps to keep a record of information needed to estimate the run-time performance of different compiler optimization sequences without the need to re-run the application again.
This technique was later referred to as speedup prediction \cite{Park2011,Kulkarni2012,Ashouri2016predictiveModeling}.

Fursin and Cohen~\cite{Fursin2007a} built an iterative and adaptive compiler framework targeting the SPEC benchmark suite using a modified version of GCC.
The authors also developed a transparent framework which reused all the compiler program analysis routines from a program transformation database to avoid duplicates in external optimization tools.

\subsection{Iterative Compilation}
\label{sec:DSE:iterative}

In computer science, an iterative method is a sequence of approximated procedures that are applied to further improve the quality of the solution of a problem.
Iterative compilation is the most commonly used exploration technique for the compiler optimization field.
Many recent works found this technique interesting and successful either (i) alone~\cite{Koseki1997,bodin1998iterative}, (ii) combined with machine learning techniques~\cite{Agakov2006,Cavazos2007,Ashouri2016Cobayn}, or (iii) combined with other search and meta-heuristics techniques such as random exploration~\cite{bodin1998iterative} and DSE techniques~\cite{nobre2015use,Martins2016TACO}.

Bodin et al.~\cite{bodin1998iterative} investigated an early path towards analysis of the applicability of iterative search techniques in program optimization.
The authors showed that iterative compilation is feasible to practice on embedded applications where the primary cost is paid back by fabrication and distribution at scale.
An embedded application usually considers with fewer parameters, so the cost of iterative compilation is worthwhile on small parameter counts.
They used profile feedback in the form of execution time and to downsample the space on restricted optimization passes.
In this work, the authors investigated unrolling, tiling, and padding parameters. Later, Other researchers, inspired by the iterative approach, explored other optimization parameters to scalar, loop-nest, and other optimizations~\cite{kisuki1999feasibility,kisuki2000iterative,KisukiK2O00,Fursin2002,Pouchet2008,Park2011,ashouri2012masterThesis,Ashouri2013VLIW,Fang2015}.

Triantafyllis et al.~\cite{Triantafyllis2003} proposed a generic optimizer that used practical iterative compilation framework called Optimization-Space Exploration (OSE).
Their approach involved using compiler writer's knowledge to prune and exclude several configuration parameters in the design space.
The OSE had function granularity and could apply different optimizations on the same code segments. They used an iterative method for selecting the next compiler optimization to be used based on the current state of the application being optimized.
A sub-classification of this approach was later named as intermediate speedup prediction \cite{Kulkarni2012,Ashouri2016predictiveModeling}.


\subsection{Non-iterative Compilation}
\label{sec:DSE:nonIterative}

Unlike iterative compilation, a non-iterative approach tries to presents a global optimization approach for a class of compiler optimization problem.
Fewer recent works are observed tackling compiler optimization problems using this method \cite{leverett1979overview}.
As we noted in Section~\ref{sec:introduction:contribution}, the driving force towards using approximation methods such as iterative compilation and machine learning was the inability of researchers to tackle the phase-ordering problem using straightforward non-iterative approaches.
Thus, this branch of compiler autotuning has suffered from further investigation.
The polyhedral compilation community has gained attention in many interesting directions and is an orthogonal approach to optimizing compilers.
We address polyhedral compilation in Section~\ref{sec:Target Domain}.

Vegdahl et al.~\cite{Vegdahl1982} have used constant-unfolding to produce code sequences that can be more compacted on a horizontal target architecture.
A constant-unfolding axiom replaces a constant by a constant expression of equal value.
The goal is to make use of constants which are hard-wired into the micro machine, replacing difficult-to-generate constants with expressions involving only hard-wired constants.

Whitfield et al.~\cite{Whitfield1997} proposed a framework that investigates the use of Gospel specifications ~\cite{whitfield1991GOSpel} for improving the performance of the code being optimized.
The authors implemented a tool called Genesis which was able to transform codes based on the Gospel specifications.
They demonstrated the benefits of the proposed framework through binary (enabling or disabling) exploration of compiler optimizations.

\section{Target Domain}
\label{sec:Target Domain}

Finding the best set of compiler optimizations to apply on a given application is heavily correlated with the type of compiler, target processor architecture, and target platform to be tuned. Research on compiler autotuning has tried to avoid generic optimizations due to this very reason.
By employing machine learning models, a framework should be adaptable based on a given application or target platform.
An optimized code segment for a given compiler or a target platform may not yield the same optimality on a different compiler or platform.
Each compiler and target platform combination have different ways of generating the binaries and executing the code segments.
Moreover, optimization techniques might be useful for a class of applications, e.g., security, scientific, etc., but they might not be for other classes.
To this end, we classify the literature based on the two aforementioned subclasses.
These are shown in Tables~\ref{tab:targetPlatform}, and ~\ref{tab:targetCompiler}.

\subsection{Target Platform}
\label{sec:Target Domain:platform}

Today, essentially all programming for desktop and HPC application is done using high-level languages (as is most programming for embedded applications).
This observation implies that since most instructions executed are the output of a compiler, an instruction set architecture is essentially a compiler target~\cite{patterson2013computer}.
For example, consider special loop instructions found in an application.
Assume that instead of decrementing by one, the compiler wanted to increment by four, or instead of branching on not equal zero, the compiler wanted to branch if the index was less than or equal to the limit.
As a result, the loop instruction may be a mismatch having different target instruction sets ISAs or architectures.
Choosing the right set of optimizations given an architecture is a necessary task.
A classification based on the type of target platform is shown in Table~\ref{tab:targetPlatform}.

\begin{table}[t]
    \centering
    \caption{A Classification Based on Target Platform}
    \hspace*{-1ex}
    \begin{tabular}{@{}|>{\footnotesize\centering\arraybackslash}m{1.2cm}|>{\footnotesize\centering\arraybackslash}m{12cm}|@{}}
        \hline
        Platform & References
        \\ \hline
         \footnotesize{Embedded Domain} &
   ~\cite{Ashouri2014bayesian,aarts1997oceans,Ashouri2013VLIW,Nobrea,Koseki1997,Cooper2002,Ashouri2016Res4ant,Fraser1999,Cavazos2006,Park2015,Martins2016TACO,Ashouri2016Cobayn,NobreRicardoLusReis2016,ashouri2012masterThesis,Martins2014,Pan2006,Dubach2007,Fursin2004,Fursin2008,Fursin2011,Ansel2014,Namolaru2010,Franke2005,Agakov2006,Kulkarni2007,Zhao2003,pallister2013identifying,nobre2015use,kelefouras2017methodology,blackmore2017automatically,bodin2016integrating,cardoso2017embedded}
        \\ \hline
         Desktop &
   ~\cite{Ashouri2017micomp, Monsifrot2002,Nobrea,Bondhugula2008a,Fursin2005,Cooper2005,Cooper2002,Yuki2012,Yuki,Park2011,Ashouri2016Res4ant,Lokuciejewski2009,Hoste2010,Kulkarni2013,Leather2009,Park2015,Cavazos2005,Lokuciejewski2009,Stephenson2006,Fursin2007a,Hoste2008,fursin2009collective,Fursin2010,Pouchet2010,Knijnenburg2003,Chen2005,Fursin2002,Cheniterativecompilationdset,Pan2006,Kulkarni2004,Li2014,Almagor2004,Purini2013,Luo2014,Stephenson2003a,Park2014,Cavazos2006a,Cavazos2004,Fursin2004,Pouchet2007,Pouchet2008,Wang2009,Stephenson2003,Cavazos2006b,Vaswani2007,Fursin2007,Fursin2008,Fursin2011,Kulkarni2012,Lokuciejewski2010,Ansel2014,Sarkar2000,Cooper1999,Haneda2005,Killian2014,Ansel2009,Queva2007,Dubach2009,Namolaru2010,Stephenson2005,Park2013,Ashouri2016predictiveModeling,Franke2005,Cavazos2007,Pan2004,Thomson2009,Mars2009,Pinkers2004,Schkufza2014,Tournavitis2009,Park2012,ogilvie2017minimizing,Agakov2006,Stock2012,Sanchez2011,kelefouras2017methodology,nobrePhase2018,CF2018ANTAREX}
        \\ \hline
         HPC Domain &
   ~\cite{Li2009,Yuki2012,Yuki,Miceli2012,NobreRicardoLusReis2016,Pouchet2007,Pouchet2008,Fursin2007,Fursin2008,Fursin2011,Ansel2014,Ansel2009,Fang2015,Schkufza2014,Tournavitis2009,Park2012,Stock2012,Sanchez2011,Tiwari2009,Ding2015,tartara2012parallel,kumar2014compiler,basu2017compiler,Ashouri2017micomp,Liu2018,gadioli2018SOCRATES,silvano2016antarex,silvano2016autotuning,silvano2015antarex,silvano2017antarex}
      \\  \hline
     \end{tabular}
    \label{tab:targetPlatform}
\end{table}

\subsubsection{Embedded Domain}
\label{sec:Target Domain:platform:embedded}

In terms of embedded computing, embedded encompasses nearly all computing that is \emph{not} considered general purpose (GP) and High Performance Computing (HPC).
Embedded processors include a large number of interesting chips: those found in cars, mobile phones, pagers, handheld consoles, appliances, and other consumer electronics~\cite{fisher2005embedded}.
Some of the more notable embedded architectures are Very Long Instruction Word (VLIW) \cite{fisher1981microcode,fishervliwnew} and the big.LITTLE heterogeneous architecture from ARM \cite{jeff2012big}.
There are many recent low-cost implementation boards with different system-on-a-chip (SoC) specifications such as Raspberry Pi \cite{upton2014raspberry} and Texas Instrument's Pandaboard \cite{instruments2012pandaboard}.
Readers can refer to the already available surveys in the field of embedded computing and FPGAs  \cite{compton2002reconfigurable,aarzen2005control,hightower2001survey}.
One of the key differences of leveraging compiler optimization techniques for the embedded domain is the trade-off between the application's code-size, performance, and power consumption.
Code-size optimization specially in VLIW architecture has been extensively investigated \cite{faraboschi2000lx,Lokuciejewski2010}.
However, due to recent advancements in the embedded SoC code-size is no longer the main issue. Thus, the focus has shifted towards the Pareto-frontiers of performance, power, and energy metrics \cite{palermo2003dse,palermo2005multi,ascia2005dse}.
Compiler optimization techniques can be exploited for this task as well \cite{fishervliwnew,Ashouri2013VLIW,NobreRicardoLusReis2016}.


Namolaru et al.~\cite{Namolaru2010} proposed a general method for systematically generating numerical features from an application.
The authors implemented their approach on top of GCC.
This method does not place any restriction on how to logically and algebraically aggregate semantical properties into numerical features; therefore, it  statistically covers all relevant information that can be collected from an application.
They used static features of MilePost GCC and MiBench to evaluate their approach on an ARC 725D embedded processor.

\subsubsection{Desktop and Workstations}
\label{sec:Target Domain:platform:desktop}

Despite the fact that the majority of the research has been done for desktops and workstations, the focus has recently shifted towards the both ends of the architectural spectrum, namely embedded and high performance computing (HPC).
In this section, we have classified those works by their experimental setup where the target platform was not either of the two ends, thus we classify them as desktops and workstation category.
This classification is shown in Table \ref{tab:targetPlatform}. 

Thomson et al. \cite{Thomson2009} proposed a clustering technique to decrease the offline training that normally supervised machine learning techniques have.
The authors achieved this goal by focusing only on those applications which best characterize the optimization space.
They leveraged the Gustafson-Kessel algorithm after applying the dimension reduction process and evaluated their clustering approach with the EEMBCv2 benchmark suite and an Intel Core 2 Duo E6750 machine.
They experimentally showed that employing technique could drastically reduce the training time.


\subsubsection{HPC Domain}
\label{sec:Target Domain:platform:HPC}

There are fundamental differences between a cluster and supercomputer.
For instance, mainframes and clusters run multiple programs concurrently and support many concurrent users versus supercomputers which focus on processing power to execute a few programs or instructions as quickly as possible and to accelerating performance to push boundaries of what hardware and software can accomplish \cite{tartara2012parallel,buyya2008market}.
However, for conciseness purposes in this survey, we have placed the recent works having used mainframes and clusters together with those having supercomputers as their experimental setup. 

Tiwari et al. \cite{Tiwari2009} proposed a scalable autotuning framework that incorporated Active Harmony's parallel search backend \cite{ctuapucs2002active} on the CHiLL compiler framework \cite{chen2008chill}.
The proposed methodology enabled the authors to explore the search space in parallel and rapidly find better transformed versions of a kernel that bring higher performance gain.

Ding et al. \cite{Ding2015} presented an autotuning framework capable of leveraging different input in two-level approach.
The framework is built upon the Petabricks language and its compiler \cite{Ansel2009}.
It uses input-aware learning technique to differentiate between inputs, clusters the space, and chooses its centroid for autotuning.
The two level approach consists of identifying a set of configurations for each class of inputs and producing a classifier to efficiently identify the best optimization to use for a new input.

Fang et al. \cite{Fang2015} proposed IODC; an iterative optimization approach which could be deployed on data centers.
They used intrinsic characteristics of data centers to overcome the well-known hurdle of running an iterative compilation method, since the technique normally requires a large number of runs per each scenario. 
IODC approach this challenge by spawning a large number of iterative compilation jobs at once to many workers and collect their performance back to a master node so that an optimal compilation policy can be rapidly found. Moreover, they evaluate their approach using a MapReduce and a compute intensive approach by, e.g., \texttt{-Ox} flags. They used the clang compiler targeting a multicore ARM processor in an ODROID board and a dual x86 desktop representative of a node in a supercomputing center.

\subsection{Target Compiler}
\label{sec:Target Domain:compiler}

In instruction-level parallelism (ILP) and super-scalar architectures \cite{wall1991limits}, parallelism mainly comes from the compiler heuristics. This insight is key to reach peak performance, energy efficiency, and lower power consumption.
Therefore, it is important to guide the compiler in order to look for the trade-off that satisfies specific objectives on a platform.
``The typical investment for a compiler back-end before maturity is measured in man-decades, and it is common to find compiler platforms with man-century investments''~\cite{fisher2004vex,fishervliwnew}.
Table \ref{tab:targetCompiler} classifies the recent literature based on the type of the compiler framework used. 

\begin{table}[t!]
    \centering
    \caption{A Classification Based on Target Compiler}
    \begin{tabular}{|>{\footnotesize\centering\arraybackslash}m{1.15cm}|>{\footnotesize\centering\arraybackslash}m{11.9cm}|}
        \hline
        Compilers & References \\ \hline

         GCC &
   ~\cite{Ashouri2014bayesian,ogilvie2017minimizing,Yuki2012,Yuki,Ashouri2016Res4ant,Leather2009,Fraser1999,Cavazos2006,Park2015,pallister2013identifying,Stephenson2006,Fursin2007a,Ashouri2016Cobayn,Hoste2008,fursin2009collective,Fursin2010,Pouchet2010,Cheniterativecompilationdset,Pan2006,Li2014,Luo2014,Park2014,Pouchet2007,Pouchet2008,Wang2009,Vaswani2007,Fursin2008,Fursin2011,Ansel2014,Haneda2005,Dubach2009,Namolaru2010,Fang2015,Franke2005,Pan2004,Thomson2009,Mars2009,Pinkers2004,Schkufza2014,Park2012,Agakov2006,Stock2012,csci1800,kumar2014compiler,kelefouras2017methodology,blackmore2017automatically,BlackmoreRE17,Liu2018,gccAutomatic2018,gongempirical,CF2018ANTAREX}
        \\ \hline
         LLVM &
   ~\cite{Ashouri2017micomp,Ashouri2013VLIW,Nobrea,Ashouri2016Res4ant,Park2015,Martins2016TACO,Ashouri2016Cobayn,NobreRicardoLusReis2016,ashouri2012masterThesis,Martins2014,Purini2013,Park2013,nobre2015use,Ashouri2016predictiveModeling,cummins2015autotuning,bodin2016integrating,nobrePhase2018,georgiou2018less,asher2017study,gongempirical}
        \\ \hline
         Intel-ICC &
   ~\cite{Bondhugula2008a,Yuki2012,Park2015,Cheniterativecompilationdset,Park2014,Pouchet2008,Killian2014,Franke2005,Schkufza2014,Park2012,Stock2012,kelefouras2017methodology,Liu2018,gongempirical}
        \\ \hline
         JIT Compiler &
   ~\cite{Hoste2010,ishizaki2015compiling,Kulkarni2013,Park2015,Cavazos2005,Stephenson2006,Cavazos2006a,Cavazos2004,Cavazos2006b,Kulkarni2012,Queva2007,Schkufza2014,Sanchez2011}
        \\ \hline
         Java Compiler &
   ~\cite{Hoste2010,Cavazos2005,ishizaki2015compiling,Cavazos2006a,Cavazos2004,Cavazos2006b,Kulkarni2012,Queva2007}
        \\ \hline
         Polyhedral Model &
   ~\cite{Bondhugula2008a,Tiwari2009,Yuki2012,Yuki,Park2015,Bondhugula2008,Pouchet2010,Pouchet2007,Pouchet2008,Park2013,wang2014energy}
        \\ \hline
         Others &
   ~\cite{Monsifrot2002,Fursin2005,Tiwari2009,Cooper2005,Cooper2002,Park2011,Ashouri2016Res4ant,Lokuciejewski2009,Kulkarni2013,Park2015,Lokuciejewski2009,Stephenson2006,Fursin2007a,Martins2016TACO,Knijnenburg2003,Chen2005,ashouri2012masterThesis,Fursin2002,Martins2014,Dubach2007,Kulkarni2004,Almagor2004,Stephenson2003a,Fursin2004,Wang2009,Stephenson2003,Fursin2007,Fursin2008,Fursin2011,Kulkarni2012,Lokuciejewski2010,Ansel2014,sarkar1997automatic,Sarkar2000,Cooper1999,Ansel2009,Queva2007,Stephenson2005,Franke2005,Cavazos2007,Tournavitis2009,Agakov2006,Sanchez2011,basu2017compiler,hallautotuning2017}
        \\ \hline
         \end{tabular}
    \label{tab:targetCompiler}
\end{table}
        
\subsubsection{GCC}
\label{sec:Target Domain:compiler:gcc}

GNU compiler collection (GCC) is the GNU compiler and toolchain project which supports various high-level languages, e.g., C, C++, etc..
``The Free Software Foundation (FSF) distributes GCC under the GNU General Public License (GNU GPL).
GCC has played an important role in the growth of free software, as both a tool and an example'' \cite{stallman2001using,stallman2003using}.
This project has been ported to various processor architectures, including most embedded systems (ARM, AMCC, and Freescale), and is able to be used on many target platforms.
Due to the wide support and open-source nature of GCC, it has been the center focus of researchers on compiler autotuning methodologies.
It is worth noting that GCC out-of-the-box does not support tuning with the phases of its internal compiler passes as its pass manager overrides predefined ordering.
However, modifying the pass manager enables tackling such compiler optimizations problem.
The GCC optimizer \footnote{https://gcc.gnu.org/onlinedocs/gcc/Optimize-Options.html} supports different predefined levels of fixed  optimization levels such as standard levels \texttt{-Ofast, -O1, -O2 and -O3}. 
Refer to the Table \ref{tab:gcco3} for the list of optimization passes inside GCC's \texttt{O3}.

 \begin{table}[t!]
 \centering
 \scriptsize
 \caption{Default Compiler Passes Inside GCC's -O3}
     \begin{tabular}{|m{13.5cm}|}
     \hline
     \multicolumn{1}{|>{\centering\arraybackslash}m{13.5cm}|}{Compiler Passes} 
     \\ \hline
     \texttt{-fauto-inc-dec -fbranch-count-reg  -fcombine-stack-adjustments  -fcompare-elim  -fcprop-registers  -fdce  -fdefer-pop  -fdelayed-branch  -fdse  -fforward-propagate  -fguess-branch-probability  -fif-conversion2  -fif-conversion  -finline-functions-called-once  -fipa-pure-const  -fipa-profile  -fipa-reference  -fmerge-constants  -fmove-loop-invariants  -freorder-blocks  -fshrink-wrap  -fsplit-wide-types  -fssa-backprop  -fssa-phiopt  -ftree-bit-ccp  -ftree-ccp  -ftree-ch  -ftree-coalesce-vars  -ftree-copy-prop  -ftree-dce  -ftree-dominator-opts  -ftree-dse  -ftree-forwprop  -ftree-fre  -ftree-phiprop  -ftree-sink  -ftree-slsr  -ftree-sra  -ftree-pta  -ftree-ter  -funit-at-a-time -fthread-jumps  -falign-functions  -falign-jumps  -falign-loops  -falign-labels  -fcaller-saves  -fcrossjumping  -fcse-follow-jumps  -fcse-skip-blocks  -fdelete-null-pointer-checks  -fdevirtualize -fdevirtualize-speculatively  -fexpensive-optimizations  -fgcse  -fgcse-lm   -fhoist-adjacent-loads  -finline-small-functions  -findirect-inlining  -fipa-cp  -fipa-cp-alignment  -fipa-bit-cp  -fipa-sra  -fipa-icf  -fisolate-erroneous-paths-dereference  -flra-remat  -foptimize-sibling-calls  -foptimize-strlen  -fpartial-inlining  -fpeephole2  -freorder-blocks-algorithm=stc  -freorder-blocks-and-partition -freorder-functions  -frerun-cse-after-loop   -fsched-interblock  -fsched-spec  -fschedule-insns  -fschedule-insns2  -fstrict-aliasing -fstrict-overflow  -ftree-builtin-call-dce  -ftree-switch-conversion -ftree-tail-merge  -fcode-hoisting  -ftree-pre  -ftree-vrp  -fipa-ra -finline-functions  -funswitch-loops  -fpredictive-commoning  -fgcse-after-reload  -ftree-loop-vectorize  -ftree-loop-distribute-patterns  -fsplit-paths -ftree-slp-vectorize  -fvect-cost-model  -ftree-partial-pre  -fpeel-loops -fipa-cp-clone} \\
     \hline
     \end{tabular}
     \label{tab:gcco3}
 \end{table}

\subsubsection{LLVM}
\label{sec:Target Domain:compiler:llvm}

The LLVM Project is a collection of modular and reusable compiler and toolchain technologies used to develop compiler front ends and back ends.
Latner and Vikram \cite{lattner2004llvm} described LLVM as ``a compiler framework designed to support transparent, lifelong program analysis and transformation for arbitrary programs, by providing high-level information to compiler transformations at compile-time, link-time, run-time, and in idle time between runs.'' 
LLVM brings many interesting features: (i) It has a common intermediate representation of code called LLVM-IR in static single assignment (SSA) form.
(ii) Its C language frontend system called \emph{clang} offers many pragma-based extensions languages.
(iii) Its backend has been ported to various architectures such as x86-64, ARM, FPGA, and even GPUs.
Recently, LLVM's community has become a vibrant research community towards porting and building new features into the different LLVM sub-modules e.g., \texttt{opt} (optimizations), \texttt{clang} (C language family front-end), and \texttt{llc} (code generator).
There are also many research papers associated with using and building LLVM \footnote{http://llvm.org/pubs/}.

Table \ref{tab:llvmo3} represents the optimization passes inside LLVM's \texttt{O3}.

 \begin{table}[t!]
 \centering
 \scriptsize
 \caption{Default Compiler Passes Inside LLVM's -O3}
     \begin{tabular}{|m{13.5cm}|}
     \hline
     \multicolumn{1}{|>{\centering\arraybackslash}m{13.5cm}|}{Compiler Passes} 
     \\ \hline
     \texttt{-tti -targetlibinfo -tbaa -scoped-noalias -assumption-cache-tracker -forceattrs -inferattrs -ipsccp -globalopt -domtree -mem2reg -deadargelim -basicaa -aa -domtree -instcombine -simplifycfg -basiccg -globals-aa -prune-eh -inline -functionattrs -argpromotion -domtree -sroa -early-cse -lazy-value-info -jump-threading -correlated-propagation -simplifycfg -basicaa -aa -domtree -instcombine -tailcallelim -simplifycfg -reassociate -domtree -loops -loop-simplify -lcssa -loop-rotate -basicaa -aa -licm -loop-unswitch -simplifycfg -basicaa -aa -domtree -instcombine -loops -scalar-evolution -loop-simplify -lcssa -indvars -aa -loop-idiom -loop-deletion -loop-unroll -basicaa -aa -mldst-motion -aa -memdep -gvn -basicaa -aa -memdep -memcpyopt -sccp -domtree -demanded-bits -bdce -basicaa -aa -instcombine -lazy-value-info -jump-threading -correlated-propagation -domtree -basicaa -aa -memdep -dse -loops -loop-simplify -lcssa -aa -licm -adce -simplifycfg -basicaa -aa -domtree -instcombine -barrier -basiccg -rpo-functionattrs -elim-avail-extern -basiccg -globals-aa -float2int -domtree -loops -loop-simplify -lcssa -loop-rotate -branch-prob -block-freq -scalar-evolution -basicaa -aa -loop-accesses -demanded-bits -loop-vectorize -instcombine -scalar-evolution -aa -slp-vectorizer -simplifycfg -basicaa -aa -domtree -instcombine -loops -loop-simplify -lcssa -scalar-evolution -loop-unroll -basicaa -aa -instcombine -loop-simplify -lcssa -aa -licm -scalar-evolution -alignment-from-assumptions -strip-dead-prototypes -globaldce -constmerge} \\
     \hline
     \end{tabular}
     \label{tab:llvmo3}
 \end{table}

\subsubsection{Intel-ICC}
\label{sec:Target Domain:compiler:icc}

Intel's propriety compiler (ICC) \footnote{https://software.intel.com/en-us/intel-compilers} provides general optimizations, e.g., \texttt{-O1, -O2, -O3} , and processor's specific optimizations depending on the target platform. Moreover, Interprocedural Optimization (IPO) is an automatic, multi-step process that allows the compiler to analyze your code to determine where you can benefit from specific optimizations.
With IPO options, you may see additional optimizations for Intel microprocessors than for non-Intel microprocessors.


\subsubsection{Just-in-time Compiler}
\label{sec:Target Domain:compiler:jit}

Just-in-time compilation (JIT), also known as dynamic translation, is a widely known technique that has been used for many decades.
``Broadly, JIT compilation includes any translation performed dynamically after a program has started execution'' \cite{aycock2003brief}. 
High-level benefits of using a JIT compiler can be summarized as:
(i) Compiled programs run faster, especially if they are compiled into a form that is directly executable on the underlying hardware.
(ii) Interpreted programs tend to be more portable.
(iii) Interpreted programs can access run-time information. 
There are many implementations of JIT compilers targeting different programming languages.
Majic, a Matlab JIT compiler \cite{almasi2000majic}, OpenJIT \cite{ogawa2000openjit} a Java JIT compiler, and IBM's JIT compiler targets the Java virtual machine \cite{suganuma2000overview}.

Sanchez et al. \cite{Sanchez2011} used SVMs to learn models to focus on autotuning the JIT compiler of IBM Testarossa and the build compilation plan. They experimentally evaluated the learned and observed that the  models outperforms out-of-the-box Testarossa on average for start-up performance, but underperforms Testarossa for throughput performance. They also generalized the learning process from learning on SPECjvm98 to DaCapo benchmark. 

\subsubsection{Java Compiler}
\label{sec:Target Domain:compiler:java}

A Java compiler is specifically designed to compile Java high-level programming code to emit platform-independent classes of Java bytecode.
Subsequently, the Java virtual machine (JVM) can load these files and (i) emit a JIT, or, (ii) interpret the bytecode into target platform machine code.
The optimizations can possibly be done in the process, technically classifying them under HIT compilation as well \cite{adl1998fast,joy2000java}.
Some of the notable research works \cite{Cavazos2004,Hoste2010} including a work tackling the phase-ordering problem have been done using Java JIT compiler \cite{Kulkarni2012}.
We have already mentioned these in the Section \ref{sec:ML_Models:Unsupervised:GA}.

\subsubsection{Polyhedral Model}
\label{sec:Target Domain:compiler:poly}

Polyhedral compilation community has recently gained attraction due to its wide appliance on the high performance computing projects \footnote{http://polyhedral.info/}.
``Polyhedral compilation encompasses the compilation techniques that rely on the representation of programs, especially those involving nested loops and arrays, thanks to parametric polyhedra or Presburger relations'' \cite{FEAUTRIER1988PARAMETRIC,WILDE1993POLYLIB}. 
Wide range of compute-intensive and research applications spend the majority of their execution time in loop-nests and this suitable for targeting high-level compiler optimizations \cite{vuduc2004HPC}.
Polyhedral compilation can address many of these challenges by efficiently transforming their loop-nest \cite{Pouchet2008,benabderrahmane2010polyhedral,Bondhugula2008a,bondhugula2008pluto,loechner1999polylib}.
We refrain from focusing more on this interesting topic as it is outside the scope of this survey.

\subsubsection{Other Compilers}
\label{sec:Target Domain:compiler:other}
        
In this survey, we focused on the classification of the more widely known compiler framework in autotuning field.
However, there are numerous other well-known compilers which worth mentioning including Cosy~\cite{alt1994cosy} and SUIF~\cite{wilson1994suif}.
We classified all  work related to use of other compiler in Table \ref{tab:targetCompiler} under others subfield.

Stanford University Intermediate Format (SUIF) compiler is a free and open-source framework implemented to support collaborative research in parallelizing and applying high-level optimizations for compilers.
It supports both Fortran and C as input languages and it can be built on top of the application.
``The kernel defines the intermediate representation, provides functions to access and manipulate the intermediate representation, and structures the interface between compiler passes'' \cite{wilson1994suif}.
It has been used in a number of recent research works \cite{Chen2005,Childers2005,Dubach2007}.

Fortran compilers include many different implementations and different ports such as Open64 \cite{developers2001open64}, GNU Fortran \cite{schulte1999interval}, XL Fortran \cite{kulkarni1997xl}, Salford Fortran 77 compiler \cite{larmouth1981fortran}, etc. and are widely used in literature \cite{sarkar1997automatic,Sarkar2000,Knijnenburg2003,Almagor2004}.

\section{Most Influential Papers}
\label{sec:influentials}

Evaluating a research work is no easy task and often involves human error.
However, in this survey, we assess and present influential work using their scientific breakthrough, novelty, and a competitive citation metric.
We discretized the process by presenting the influential papers by their corresponding topic and elaborate more on their proposed approach.
Some had effects on their succeeding work, and this was taken into consideration as well.

\subsection{Breakthroughs by topic}
\label{sec:influentials:byNovelty}
The following paragraphs highlight novel research in the areas of initial introduction of learning methods with compiler optimizations, genetic algorithms, phase ordering, iterative compilation, dynamic and hybrid features, creating optimization groups with Bayesian learners, and clustering of optimizations to tackle the phase-ordering problem.

\paragraph{Introducing Learning Methods}
Leverett et al. and Vegdahl et al. \cite{leverett1979overview,Vegdahl1982} were the first to perform non-iterative optimization without leverage machine learning.
\cite{Whitfield1990} extended this work by proposing intelligent ordering of a subset of compiler optimizations.
\cite{Whitfield1997} continued their phase-ordering work with a formal language, Gospel, which could be used to automatically generate transformations.
The first usage of machine learning techniques arrived with \cite{Koseki1997} and their work with predicting the optimal unroll size for nested loops.
\cite{cavazos1998} was the first to use machine learning techniques to construct flexible instruction schedules, paving the way for continued efforts leveraging ML techniques.

\paragraph{Genetic Algorithms}
Cooper et al. \cite{Cooper1999} expanded machine learning efforts with optimization selection using genetic algorithms with iterative compilation.
\cite{Cooper2002} expanded their prior work by switching to adaptive compilation and one of the earliest works creating an adaptive compilation framework.
Predictive modeling was first introduced by \cite{Triantafyllis2003} where they applied iterative compilation of the SPEC benchmarks.
\cite{Knijnenburg2003} proposed iterative compilation to select tile and unroll factors using genetic algorithms, random sampling, and simulated annealing.
They were able to show that their method worked on many different architectures.

\paragraph{Phase Ordering}
Kulkarni et al. \cite{Kulkarni2004} was one of the first to propose solving the phase-ordering problem using machine learning by combining iterative compilation and meta-heuristics.
\cite{Kulkarni2012} tackled the phase-ordering problem within the JIKES Java virtual machine.
They leveraged static features fed into a neural network generated with NEAT to construct good optimization sequence orders.

\paragraph{Iterative compilation}
Bodin et al. \cite{bodin1998iterative} proposed an intelligent iterative compilation method which explored less than 2\% of the total space in a non-linear search space.
\cite{Agakov2006} used Markov chains to focus iterative optimization using static features.
Using a relatively small exhaustive search space for learning ($14^5$) and a large test space for testing ($80^{20}$), they were able to achieve up to 40\% speedup.

\paragraph{Dynamic and Hybrid Features}
The first use of dynamic features for learning was introduced by Cavazos et al. \cite{Cavazos2007};
they showed that using dynamic features for learning outperformed the use of static features.
The advancement of multivariate (static, dynamic, hybrid) feature selection and learning algorithms paved the way for tournament predictors introduced by \cite{Park2011}.

\paragraph{Practical and Collaborative Autotuning}
MILEPOST GCC \cite{Fursin2008,Fursin2011} was the first attempt to make a practical on-the-fly machine-learning based compiler combined with an infrastructure targeted to autotuning and crowdsourcing scenarios.  
It has been used in practice and revealed many issues yet to be tackled by researchers including (1) reproducibility of empirical results collected with multiple users and (2) problems with metadata, data representation, models, and massive datasets \cite{fursin2015collective,fursin2016collective}.

\paragraph{Hybrid Characterization and Bayesian Learners}
Massive dataset analysis on over 1000 benchmarks was performed by \cite{Cheniterativecompilationdset,chen2012deconstructing}.
They proposed optimization sequence groups (beyond traditional compilers' \texttt{-O3} baseline) that are, on average, beneficial to use on the applications in their dataset.
Most recently, Ashouri et al. \cite{Ashouri2014bayesian,Ashouri2016Cobayn} used the output of the passes suggested by \cite{chen2012deconstructing} to construct a Bayesian network to identify the best compiler flags for a given application using static, dynamic, and hybrid features.
The Bayesian network generated optimization sequences resulted in application performance outperforming existing models.

\paragraph{Optimization Clustering and Full-sequence Predictors.  }
\major{Ashouri et al. \cite{Ashouri2017micomp} introduced MiCOMP framework to cluster all the LLVM's \texttt{-O3} optimization passes into optimization sub sequences and introduce the first full-sequence speedup predictor for the phase-ordering problem. The authors leveraged a recommender systems  \cite{ricci2011introduction} approach to defined exploration policy using dynamic information stored in the pair-wised optimizations across the training applications to outperform the state-of-the-art ranking approach. They show MiCOMP can achieve over 90\% of the available speedup and outperform \texttt{-O3} using just a few predictions.}

\subsection{Breakthroughs by Performance}
\label{sec:influentials:byPerformance}

\major{This section provides a brief quantitative comparison between the proposed approaches regarding their reported results. Evaluating papers using their performance is a hard task and often involves comparison errors \cite{hoefler2015scientific}, i.e., using absolute vs. relative speedup values, averaging using different techniques, etc..
To this end, we look into papers which explicitly provided comparison of their proposed approach against their baseline or the state-of-the-art methods.} 

\subsubsection{Iterative Compilation} 
\major{\emph{Random Iterative compilation} (RIC) is known to achieve good results when compiling long running applications \cite{bodin1998iterative}.
However, the approach is expensive and should be combined with an intelligent search algorithms, such as machine learning techniques \cite{Ashouri2017micomp,Ashouri2016Cobayn,Agakov2006,bodin1998iterative}.
Early work in iterative compilation methods \cite{aarts1997oceans,bodin1998iterative} involved the exploration of non-linear transformation spaces and finding the fastest execution time in a fixed number of evaluations targeted to embedded domain. In embedded domains, the cost of performing iterative compilation is amortized over the deployment of a large number of devices. Researchers used profile information in the form of execution time, searched a large but restricted subset of transformations to find good results, and often achieve large speedups by exploring only a small fraction of the optimization space. Other work outperformed these approaches by introducing more efficient search algorithms on the space in addition to considering a more extensive optimization space with different unroll factors and tile sizes \cite{Knijnenburg2003}. 
Mpeis et al. \cite{MpeisPL15iterativeMobile} implemented an iterative compilation approach targeted to mobile devices to further optimize the Java bytecode of day-to-day running application when the device was not being used. The authors observed on average a 57\% performance improvement using a benchmark with minimal slowdowns.
Other major works tried to incorporate iterative compilation methodology with machine learning techniques and we will discuss this category next.}

\subsubsection{Machine Learning Techniques}
\paragraph{The Selection Problem}
\major{Agakov et al. \cite{Agakov2006} introduced a machine learning approach to focus on iterative compilation.
Their approach using static features and Markov oracle led to 22\% and 27\% performance improvement on a Texas instrument and an AMD architecture.
Later, Cavazos et al. \cite{Cavazos2007} outperformed the previous work by 7\% using a dynamic feature characterization method.
The authors believed that using static features of an application although is good enough for embedded multimedia applications but cannot perform well on a large scale general purpose applications.
The use of predictive modeling approaches went on by other authors \cite{Fursin2008,Hoste2008,fursin2010collective}.
Park et al. \cite{Park2011} later introduced a novel approach using tournament predictors (see Section \ref{sec:predication_classes:tournametIntermdiatePrediction}), by which they outperformed existing methods by around 6\%.
The authors went on to propose to use graph kernels \cite{Park2013} (see Section \ref{sec:predication_classes:tournametIntermdiatePrediction}) as a means of improving existing prediction models.
Ashouri et al. introduced COBAYN \cite{Ashouri2016Cobayn}, which used hybrid features and Bayesian network learners.
The authors showed that this approach can outperform existing methods \cite{Agakov2006,Park2015} by around 11\%. }

\paragraph{The Phase-ordering Problem}
\major{Cooper et al. \cite{Cooper2002} proposed a search mechanism for finding good ordering of phases using genetic algorithm.
The approach relied upon an adaptive compilation flow that could gain up to 49\% speedup on the given benchmark.
Later, Kulkarni et al. \cite{Kulkarni2004} proposed yet another genetic algorithm which was able to drastically reduce the search time up to 65\%.
The first intermediate speedup predictor was later proposed by Kulkarni and Cavazos \cite{Kulkarni2012}.
The authors used the characterization of the current state of the code and NEAT (refer to Section \ref{sec:ML_Models:Unsupervised:GA}) and gained up to 20\% speedup on applications using Java Jikes compiler.
Ashouri et al. proposed MiCOMP \cite{Ashouri2017micomp}, which uses subsequences derived from LLVM's \texttt{-O3} and a full-sequence speedup predictor.
The authors showed that using their approach they could outperform existing approaches \cite{Kulkarni2012,Ashouri2016predictiveModeling} by 5\% and 11\%, respectively.}

\subsubsection{Multi objective Optimization}
\major{In this subsection, we briefly discuss the quantitative results of investigating the major optimization objectives drawn in the literature.
These objectives are code size, area, power metrics, and performance metrics.}

\paragraph{Code Size and Area.}
\major{Early works were targeting code-size reduction as an optimization objective.
Cooper et al. \cite{Cooper1999} introduced a Genetic algorithm (GA) method to search for reduced code size and found it had an advantage of 14.5 \% against a default fixed sequence.
These works were followed by a series of others mostly tackling the code size reduction in VLIW embedded architecture \cite{fisher2005embedded,ascia2005dse,wong2008vliw,Ashouri2013VLIW}.
However, through the advancement of storage systems specifically in the embedded domain, this issue has become far less of a concern and often has been neglected in the recent literature.}

\paragraph{Performance and Intensity.}
\major{Roofline model \cite{williams2009roofline} relates processor performance to off-chip memory traffic, thus provides a theoretical upper bound for operational intensity and attainable performance in modern architectures.
Certain works have used the notion to model their optimization approaches and find a Pareto curve to satisfy both objectives \cite{Hoste2008, Ashouri2013VLIW,deb2002fast,ashouri2012masterThesis,palermo2005multi,zaccaria2010multicube,ascia2005dse}.
Hoste and Eeckhout \cite{Hoste2008} tackled iterative compilation on an Intel Pentium machine and observed the feasibility of their approach together with a multi-objective search algorithm called multiple populations which finds Pareto curves within the different population of an optimization space.
They observed up to 37\% speedup when they used their evaluation metric.
Ashouri et al. \cite{Ashouri2013VLIW} applied roofline model to a customized VLIW architecture and formed four clusters to satisfy a multi-objective proposed scenario.
They observed reaching a speedup of up to 23\% in execution time by using those compiler optimizations found in their method that contributed to the satisfaction of the objectives.} 

\paragraph{Power.}
\major{Except for a few recent works \cite{NobreRicardoLusReis2016,wang2014energy}, this objective has been mostly investigated at architectural and system-level \cite{mcpat,palermo2005multi,ascia2005dse,ayala2007energy}.
For this reasons, it is outside the scope of this survey. }

\subsection{Citation Metric}
\label{sec:influentials:byCitation}

In this section, we show the top 16 most-cited papers among more than 200 papers we elaborated \footnote{Data has been extracted from Google Scholar on July 2018 and they subject to change.}.
in Table \ref{tab:citation}, we present their citation count and the average citation per year (ACPY) with a few keywords representing their methodology we already covered in this survey.

\begin{table}[!t]
\centering
\footnotesize
\caption{A Classification of Top 16 Influential Papers By Their Citation Count}
\label{tab:citation}
\hspace*{-1ex}
\begin{tabular}{ | >{\footnotesize}M{0.2cm} | >{\footnotesize}m{3.7cm} | >{\footnotesize}M{0.8cm} | >{\footnotesize}M{0.5cm} |  >{\scriptsize}m{6.7cm}| }
\hline
	No &  Reference & Citation Count & ACPY &  Keywords \\ \hline
	1 & Agakov et al. [2006] \cite{Agakov2006} & 377 & 33 & iterative compilation, static features, Markov chain oracle, PCA \\ \hline
	2 & Cooper et al. [1999] \cite{Cooper1999} & 322 & 17 & genetic algorithm, iterative compilation, reduced code-size \\ \hline
	3 & Triantafyllis et al. [2003] \cite{Triantafyllis2003} & 277 & 19 & iterative compilation, SPEC, predictive modeling \\ \hline
    4 & Stephenson et al. [2003] \cite{Stephenson2003} & 274 & 19 &   iterative compilation, genetic programming, metaheuristics \\ \hline
    5 & Fursin et al. [2008,2011]\cite{Fursin2008,Fursin2011} & 251 & 22 &  MilePost GCC, self-tuning, crowdsourcing, Iterative compilation \\ \hline
	6 & Cooper et al. [2002] \cite{Cooper2002} & 242 & 15 &  adaptive compilation, biased random search, sequence predictor \\ \hline
    7 & Knijnenburg et al. [2003] \cite{Knijnenburg2003} & 238 & 16 &  iterative compilation, unrolling factor, architecture-independent \\ \hline
	8 & Tournavitis et al. [2009] \cite{Tournavitis2009} & 234 & 27 & static-analysis, profile-driven parallelism, NAS, SPEC \\ \hline
	9 & Cavazos et al. [2007] \cite{Cavazos2007} & 225 & 20 &  iterative compilation, dynamic characterization, sequence predictor \\ \hline
    10 & Tiwari et al. [2009] \cite{Tiwari2009} & 200 & 25 & CHiLL framework, iterative compilation,  sequence predictor \\ \hline
    11 & Almagor et al. [2004] \cite{Almagor2004} & 196 &  14 &  adaptive compilation, compiler sequence predictor, SPARC  \\ \hline
	12 & Monsifrot et al. [2002] \cite{Monsifrot2002} & 175 &  11 & decision trees, boosting, abstract loop representation \\ \hline
	13 & Stephenson et al. [2005]  \cite{Stephenson2005} & 170 & 13 & supervised learning, unrolling factor, multiclass classification \\ \hline
	14 & Pan et al. [2006] \cite{Pan2006} & 161 & 13 & combined elimination, iterative compilation, SPEC \\ \hline
	15 & Bodin et al. [1998] \cite{bodin1998iterative} & 151 & 7 & iterative compilation, multi-objective exploration \\ \hline
	16 & Hoste et al. [2008] \cite{Hoste2008} & 134 & 13 & iterative compilation, metaheuristic, genetic algorithm \\ \hline
\end{tabular}
\end{table}

\section{Discussion \& Conclusion}
\label{sec:conclusion}

In the coming decades, research on compilation techniques and code optimization will play a key role in tackling and addressing various challenges of computer science and the high performance computing field.
This includes auto-parallelization, security, exploiting multi and many-core processors, reliability, reproducibility, and energy efficiency.
Using compiler optimizations to exploit large-scale parallelism available on architectures and power-aware hardware is an essential task.
Additionally, machine learning is becoming more powerful by leveraging deep learning to find and construct heuristics.
These complex learners allow automated systems to efficiently perform these task with minimal programmer effort.

\minor{Additionally, research on collaborating tuning methodologies have gained attention by the introduction of Collective Knowledge framework (CK) \cite{fursin2018collective,Fursin2011,memon2013crowdtuning,fursin2016collective}. CK is a cross-platform open research SDK developed in collaboration with academic and industrial partners to share artifacts as reusable and customizable components with a unified, portable, and customizable experimental work flows. Researchers can automate the tuning process across diverse hardware and environments. Crowdsourcing and reproducing experiments across platforms provided by other researchers would also be feasible. CK is now used as an official platform to support open ACM ReQuEST tournaments on reproducible and Pareto-efficient software/hardware co-design of artificial intelligence, deep learning and other emerging workloads \cite{CK-ReQuest}. }

\major{By the advent of \emph{Deep Learning}, i.e., deep neural networks (DNN) \cite{schmidhuber2015deep}, we are witnessing more and more applications of such techniques, i.e., speech recognition \cite{hinton2012deep}, image classification \cite{krizhevsky2012imagenet}, etc..
Deep learning algorithms extract high-level and complex abstractions from data through a hierarchical learning process and have shown effectiveness in many applications.
DNNs normally require large data to train, thus adapting new benchmarks and large data sets will play an important role in achieving the potential benefits of using DNNs.
Recently, we have seen benchmark synthesizers which are able to generate synthetic programs for use in machine learning-based performance autotuning.
Genesis \cite{chiu2015genesis} and CLgen \cite{cummins2017synthesizing} are such examples that bring diversity and control to the generation of new benchmarks to the user.}

In this survey, we have synthesized the research work on compiler autotuning using machine learning by showing the broad spectrum of the use of machine learning techniques and their key research ideas and applications.
We surveyed research works at different levels of abstraction, application characterization techniques, algorithm and machine learning models, prediction types, space exploration, target domains, and influence.
We discussed both major problems of compiler autotuning, namely the selection and the phase-ordering problem along with the benchmark suits proposed to evaluate them.
We hope this article will be beneficial to computer architects, researchers, and application developers and inspires novel ideas and opens promising research avenues.



{\small	
\bibliographystyle{ACM-Reference-Format-Journals}
\bibliography{acmsmall-sample-bibfile}}




\end{document}